\documentclass{aa}

  \usepackage{graphicx}
  \usepackage[varg]{txfonts}
  \usepackage{lscape}
  \usepackage{natbib}
  \usepackage{ulem}
  \bibpunct{(}{)}{;}{a}{}{,}% to follow the A&A style

\begin{document}

     \title{Stability of planets in triple star systems}
  %   \subtitle{XXXXX}

     \author{F. Busetti
            \inst{1}
            \and
            H. Beust\inst{2}\
            \and
            C. Harley\inst{1}\
            }

     \institute{Univ. of the Witwatersrand, CSAM, Private Bag 3, 2050-Johannesburg, South Africa\\
                \email{francobusetti@iafrica.com}
           \and
               Univ. Grenoble Alpes, CNRS, IPAG, F-38000 Grenoble, France\\
               \email{herve.beust@univ-grenoble-alpes.fr}
  %             \thanks{}
               }

     \date{Received 27 March 2018 / Accepted 27 August 2018}

  % \abstract{}{}{}{}{}
  % 5 {} token are mandatory

\abstract
     {Numerous theoretical studies of the stellar dynamics of triple systems have been carried out, but fewer purely empirical studies that have addressed planetary orbits within these systems. Most of these empirical studies have been for coplanar orbits and with a limited number of orbital parameters.}
     {Our objective is to provide a more generalized empirical mapping of the regions of planetary stability in triples by: considering both prograde and retrograde motion of planets and the outer star; investigating highly-inclined orbits of the outer star; extending the parameters used to all relevant orbital elements of the triple’s stars and expanding these elements and mass ratios to wider ranges that will accommodate recent and possibly future observational discoveries.}
     {Using \textit{N}-body simulations we integrated numerically the various four-body configurations over the parameter space, using a symplectic integrator designed specifically for the integration of hierarchical multiple stellar systems. The triples were then reduced to binaries and the integrations repeated, to highlight the differences between these two types of system.}
     {This established the regions of secular stability and resulted in 24 semi-empirical models describing the stability bounds for planets in each type of triple orbital configuration. The results were then compared with the observational extremes discovered to date, to identify regions that may contain undiscovered planets.}
     {}

     \maketitle

  \section{Introduction}
      A large portion of stellar systems are composed of multiple stars. Single stars account for approximately half, binaries a third, and triples a twelfth \citep{toko2014a,toko2014b}.
Over more than two hundred years there have been numerous theoretical studies of the three-body problem and its restricted form. A few recent theoretical approaches to the stellar dynamics of triple systems, some incorporating numerical checks of their results, include \citet{ford2000,mard2001,hame2015} and \citet{corr2016}. Theoretical approaches can, however, have limitations in accurately representing some actual systems.

     Numerical studies are a powerful tool with which to map the often discontinuous stability boundaries of systems without any restriction on their physical characteristics and configurations. Also, a number of studies have shown that, for systems that contain more than one planetary body, the orbits proposed initially were simply not dynamically feasible, which has promoted dynamical analyses as an integral part of the exoplanet discovery process \citep{horn2012}.

      A hierarchical triple stellar system consists of three bodies organized in two nested orbits --- one central binary and a third star orbiting the center of mass of this binary at a larger distance. We shall call the two bodies constituting the binary stars 1 and 2, and the outer body star 3. We note that we have made no assumptions about the relative masses of the three stars. In many systems, star 3 is less massive than the central binary, but there are examples of systems with inverted mass ratios, such as HD 181068, for which star 3 is heavier than the binary \citep{dere2011,bork2013}.

      There are four possible types of planetary orbits in such a system (see Fig. \ref{Fig1}):

      - S1, S2, and S3 orbits comprise planets that orbit one of the individual stars (stars 1, 2 and 3 respectively) while being perturbed by the other stars. S1 and S2 orbits are not distinguishable and will be treated as a single S1 type;

      - P1 orbits comprise circumbinary planets that orbit the central binary while being perturbed by the outer third star; and

      - P2 orbits comprise circumtriple planets that orbit the whole system beyond star 3.

      Planets have been found in 32 triple systems to date. Of these, S3 orbits have been found in 29 systems, S1 orbits in two and a P1 orbit in one. No planet in a circumtriple P2 orbit has yet been discovered.

       Numerical studies of planets in multiple systems have focused on binaries. For example, in their key paper, \citet{holm1999} examined S-type and P-type planetary orbits in eccentric, coplanar binary systems and derived regression equations describing the stability bounds in terms of mass ratios and eccentricities. Eccentric and inclined (up to 50$^\circ$) P-type orbits, for equal-mass binaries, are studied by \citet{pila2003}. Work on S-type and P-type orbits is done by \citet{musi2005} for circular, coplanar orbits, defining the regions of stability as a function of the semi-major axis ratio between the star and planet and the mass ratio. \citet{mudr2006} also analyze the numerical results of \citet{holm1999} for S-type orbits in coplanar systems and investigated the instability boundary at a higher resolution. A binary study that addressed a full range of planetary inclinations, for eccentric P-type orbits, was by \citet{dool2011}, which described stability as a function of mass ratios and eccentricities.

      One study that addressed retrograde planets in binary, coplanar, circular, S-type orbits, was by \citet{mora2012}, while \citet{gupp2017} looked at S-type orbits, including retrograde ones, in a compact binary system, for various inclinations, eccentricities and orientation angles. Although none have yet been confirmed, planets in retrograde orbits within and around a triple system should be possible, similar to HAT-P-7b in its binary system.

      One study of the four-body problem of planets within a triple system was by \citet{verr2007}. This study was confined to prograde, coplanar P-type orbits and derived regression relationships analogous to those of \citet{holm1999}.

      In a more general analysis, in addition to retrograde planetary orbits it is also necessary to include highly-inclined orbits of the outer star, since the resulting stellar Kozai resonance sculpts the geometry of the stable planetary region \citep{koza1962,ford2000}. It is also useful to consider retrograde motion of the outer star. For the 22 triple systems where mutual inclinations have been established, the mean is a high 63$^\circ$ and four of these systems are retrograde \citep{toko2017}. One analytical study that addressed retrograde stellar orbits is by \citet{farag2010}. They address two limiting cases, the inner restricted problem, where one inner body (i.e., our star 2) has no mass, and the outer restricted problem, where the outer body (star 3) has negligible mass, describing the possible motions of particles in each regime and providing an expression for the boundary between the regimes.

      We extended the number of parameters used in previous work to all the orbital elements of a triple’s stars and widened the ranges of both these and the relevant mass ratios, to encompass recent observations as well as potential future discoveries. For example, systems with highly inverted mass ratios, such as HD 181068, have been discovered but not yet modeled.

      The objective of this work was to provide a broader mapping of the regions of planetary stability in generalized triples. This resulted in 24 semi-empirical models describing the stability bounds for planets in each possible combination of orbital motions in triples and for a range of stellar orbital elements and mass ratios.

      \section{Methods}
      \subsection{Selection of parameter space}
\begin{figure}[htbp]
     \centering
     \includegraphics[width=\hsize]{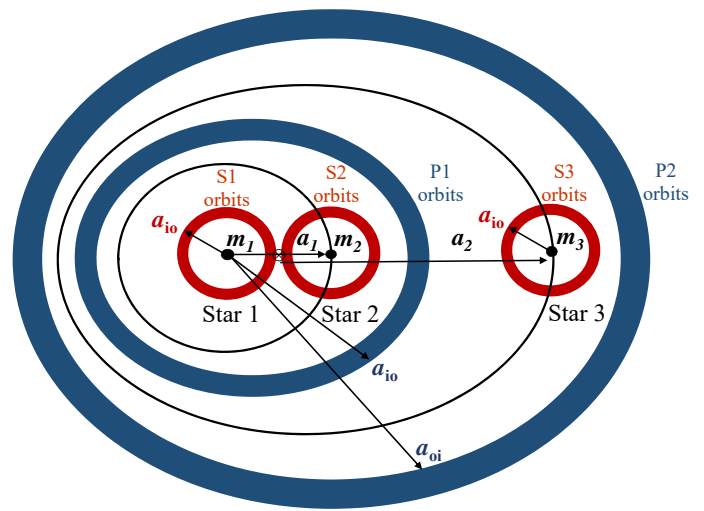}
     \caption{Triple system with S-type and P-type orbits.}
     \label{Fig1}
     \end{figure}

      Our aim was to determine the orbital stability of both the host stars and their planets, where collisions between the stars or the ejection of one of the three stars (typically the least massive body) does not occur over secular timescales that are very long compared to the orbital periods. Our treatment of secular perturbations was based on classical Newtonian dynamics and assumed that all three bodies are point masses that do not interact other than through gravitation and do not evolve in any way. Relativistic and tidal effects, as well as spin, were not addressed. The triple system’s nomenclature is shown in the schematic illustration in Fig. \ref{Fig1}.

      There are two S-type orbits, as the S1 and S2 orbits are equivalent, and two P-type orbits, one circumbinary and one circumtriple. The stable region for circumbinary (P1) orbits will have an inner and outer bound, while that for circumtriple (P2) orbits will have an inner bound only. Circumstellar (S1 and S3) regions of stability will have their outer edges bounded. The objective is to determine the outer bound $a_{\mathrm{io}}$ of the stable regions for S1, S3, and P1 orbits and the inner bound $a_{\mathrm{oi}}$ of the P2 orbits. The following subsections list the selected parameter ranges for the stellar configurations.
      \medskip
      \\2.1.1. Semi-major axis ratio $a_{2}/a_{1}$
      \\The lower limit was selected to be just inside the stability criterion for triples as given by \citet{mard2001}. The smallest ratio used was approximately three, lower than the smallest ratio of 4.1 found in the \citet{eggl2008} catalog of 285 triples and well above the $a_{2}/a_{1}\gg2$ requirement of the HJS symplectic integrator used. An upper limit of 100 was used; most of the semi-major axis ratios for triple systems fall within this range.
      \medskip
      \\2.1.2 Inner mass ratio $\mu_{1}= m_{2}/(m_{1}+m_{2})$
      \\The range for this ratio is between zero and one. However, only the range $0<\mu_{1}<0.5$ needs to be studied, as mass ratios of $\mu_{1}$ and $(1-\mu_{1})$ are equivalent, other than for a 180$^\circ$ change in the longitude of the ascending node. An upper limit of one and lower limits of 0.001 and 0.1 were used for P and S orbits respectively.
      \medskip
      \\2.1.3. Outer mass ratio $\mu_{2}= m_{3}/(m_{1}+m_{2})$
      \\Outer mass ratio definitions vary widely. In the \citet{toko2014a} survey, the largest mass ratio (defined there as $m_{3}/m_{1})$ found was 8.9 (for HIP 29860). However, 94\% of mass ratios are below two and 99\% are below five. Taking an upper limit of five and assuming that the masses of the binary pair are broadly comparable, the equivalent value for our ratio definition is 2.5. A mass ratio greater than one implies an inverted system, where the outermost star is more massive than the aggregate inner binary. An example is the triple system HD 181068, which has a mass ratio of $\sim$1.7. The lowest mass ratio found in the survey was 0.07. We therefore used ratios ranging from  0.2 or 0.001, for P and S orbits respectively, up to 2.5.
      \medskip
      \\2.1.4. Inner and outer eccentricities $e_{1}$ and $e_{2}$
      \\The only data on inner and outer stellar eccentricities in triples is from \citet{ster2002}. The mean eccentricity of all orbits is high at 0.39 and in most (70\%) of the systems the inner orbit is more eccentric than the outer orbit. The difference in eccentricities within these systems ranges from effectively zero to 0.57. A study of 222 Kepler triples finds outer eccentricities spanning the full range, with a broad peak in the middle of the range and an unexplained narrow peak near $e_{2}\simeq0.28$ \citep{bork2015}. The integrations therefore covered eccentricities in both inner and outer orbits ranging from zero to 0.7--0.9 to cover fully this parameter range, with the higher limit required when Kozai resonance occurs.
      \medskip
      \\2.1.5. Inner and outer inclinations $i_{1}$ and $i_{2}$
      \\The inner inclination $i_{1}$ was always set to zero, so the outer inclination $i_{2}$ was the mutual inclination $i_{2}-i_{1}$ of the outer orbit relative to the inner orbit. For the low-inclination simulations, small ranges of mutual inclination were selected, of 0$^\circ$--60$^\circ$ for prograde orbits and 120$^\circ$--180$^\circ$ for retrograde orbits. In the high-inclination integrations the full range of 0$^\circ$--180$^\circ$ was used. \citet{bork2015} found that the distribution of mutual inclinations for 62 Kepler triples had a large peak at 0$^\circ$--5$^\circ$, indicating close-to-coplanar configurations, with a significant 38\% portion of the systems in a secondary peak centred at 40$^\circ$, suggesting Kozai effects.
      \medskip
      \\2.1.6. Other orbital elements
      \\The outer star’s longitude of ascending node $\Omega_{2}$ and argument of periapsis $\omega_{2}$ were both varied from 0$^\circ$ to 360$^\circ$ while those for the inner star were set to zero.
       \medskip
       \\2.1.7. Test particles
       \\Planets were represented by massless test particles. Their initial orbital elements ranged between selected lower and upper limits --- for eccentricity these were 0--0.9, for inclination 0$^\circ$--90$^\circ$ or 90$^\circ$--180$^\circ$ for prograde or retrograde orbits respectively and for $\omega$, $\Omega$ and mean anomaly $M$ they were 0$^\circ$--360$^\circ$. Most initial orbital elements were randomly generated from uniform distributions. However, for inclination $i_{2}$ the direction of angular momentum should be uniform over the celestial sphere, which requires a uniform distribution in $\cos i_{2}$ between [-1,1].

       A log scale was used for semi-major axes. The particles' semi-major axes ranged between 0.02 AU and 0.9$a_{1}$ for S1 orbits and 0.02 AU and $fa_{2}$ for S3 orbits, where $f$ varied between 0.07 and 0.60 depending on the stellar configuration. For P1 and P2 orbits the lower limit was 0.01 AU while the outer limit was set conservatively at twice the distance at which the 5:1 mean motion resonance occurred. The planet orbiting closest to its central body is PSR 1719-14 b, at 0.0044 AU, while Kepler-42c has the closest orbit to a “normal” star, at 0.006 AU. The distance at which a test particle was stopped as being too close to the central body was set at 0.002 AU. The minimum semi-major axis used must be larger than this, and so 0.01--0.02 AU was selected for all orbit types. The furthest planet from its host star yet discovered is HIP 77900 b, at 3\,200 AU. The distance at which a test particle was assumed to have escaped from the central body was set at 10\,000 AU.

      \subsection{Numerical methods}
      %\medskip
      2.2.1. Integrator
      \\We used the symplectic HJS (Hierarchical Jacobi Symplectic) integrator developed by \citet{beus2003} from the Swift code \citep{levi1994,levi2013}. It is designed specifically to allow the integration of multiple stellar systems and to handle hierarchical systems of any number and structure, provided the hierarchy is preserved along the integration. The HJS algorithm has proved itself a valuable tool in many studies of hierarchical systems. Most triples are hierarchical, simply because if they were not, the system is likely to be unstable and fragment into a binary and an ejected third star.
      \medskip
      \\2.2.2. Computational parameters
      \\An integration time step of $1/20$ of the orbital period of the inner binary was normally used in the integrations, and varied to verify that results were not affected by numerical errors. Symplectic integration schemes usually ensure energy conservation with $10^{-6}\mbox{--}10^{-8}$ relative accuracy using this time step size \citep{levi1994,beus2003,verr2007}. We found an overall fractional change in the system energy $\Delta{E}/E_{0}$ of $\sim$$10^{-7}$ over a $10^5$ yr integration, although this varied widely depending on the triple configuration being integrated. Orbital instabilities tended to occur quickly -- if an orbit did not become unstable in as little as 10--100 yr, it was usually stable up to $10^5$ yr. The fact that the stellar systems we investigated were largely compact was helpful, as $10^5$ yr was often equivalent to many hundreds of thousands of orbits of the outer star.

      The secular evolution of orbital parameters was sampled for each of the configurations used, and an integration time of $10^5$ yr was found to be sufficiently long in most cases. Each batch of integrations included at least one run on the least compact configuration with an integration time of $10^6$ yr or $10^7$ yr,  to confirm this.

      The number of planetary test particles used varied from 1\,000 to 10\,000, with higher numbers being used when the rate of instability and hence removal of test particles was high. For coplanar configurations, the initial test particle cloud took the form of a disk aligned with the plane of the inner binary, while for configurations where the outer star had a high inclination the test particle disk was aligned with the invariable plane.
A procedure for automatically identifying the edges of the test particle cloud at the end of an integration and measuring its semi-major axis was incorporated into the HJS algorithm.

      \section{Results and discussion}
      \subsection{Orbit types P1 and P2}
      \subsubsection{Results}
      For inner (P1) and outer (P2) orbits, the respective outer and inner edges of the cleared area of the test particle cloud were standardized by taking them as ratios of the semi-major axis of the outer star, $a_{2}$. These standardized dependent variables $a_{\mathrm{io}}$/$a_{\mathrm{2}}$ and $a_{\mathrm{oi}}$/$a_{\mathrm{2}}$ delineate the bounds of the stable regions and are termed the inner and outer critical semi-major axis ratios respectively.

      The mean critical semi-major axis ratio for these inner and outer orbits, for the eight possible combinations of orbital motion, are shown in Table \ref{table:1}. These are the average ratios found over all the combinations used in the parameter space. For P1 planetary orbits the mean critical ratio is materially different (36\%) for prograde and retrograde planetary orbits. Both these ratios are slightly ($\sim$2\%) larger for retrograde stellar orbits. For P2 orbits the mean critical ratio is similarly (-33\%) different for prograde and retrograde planetary motions. For retrograde stellar orbits these ratios are slightly ($\sim$3\%) smaller. The difference made by the orbital direction of the outer star is small for both prograde and retrograde planetary orbits, but is statistically meaningful for inner retrograde planetary orbits and outer prograde planetary orbits.

      \begin{table}[htbp]
      \centering
      \caption{Mean critical semi-major axis ratios for all combinations of orbital motions in P1 and P2 orbits.}
      \label{table:1}
      \setlength{\tabcolsep}{4pt} % Default value: 6pt
      \begin{tabular}{*8{c}}
  \hline
  Orbit & Critical & \multicolumn{2}{c}{Motions${}^{1}$} & \multicolumn{4}{c}{Mean
  critical ratio} \\
  type & ratio & Star 2 & Planet & Min & {Mean} & $\sigma$ & Max \\ \hline
  %\multirow{4}{*}{P1 & \textit{a${}_\mathrm{io}$/a}${}_{2}$\textit{}} & P & P & 0.131 & 0.383 & 0.147 & 0.892 \\
  P1 & ${a_{\mathrm{io}}/{a_{2}}}$ & P & P & 0.131 & {0.383} & 0.147 & 0.892 \\
   & \textit{} & ~ & R & 0.113 & {0.519} & 0.108 & 0.828 \\ \cline{3-8}
   & \textit{} & R & P & 0.110 & {0.385} & 0.077 & 0.897 \\
   & \textit{} & ~ & R & 0.112 & {0.537} & 0.132 & 0.857 \\ \hline
  P2 & ${a_{\mathrm{oi}}/{a_{2}}}$ & P & P & 1.253 & {2.936} & 1.449 & 5.197 \\
   & \textit{} & ~ & R & 1.184 & {1.976} & 0.507 & 4.916 \\ \cline{3-8}
   & \textit{} & R & P & 0.924 & {2.773} & 0.891 & 5.384 \\
   & \textit{} & ~ & R & 0.591 & {1.960} & 0.571 & 4.957 \\ \hline
  \multicolumn{8}{p{2in}}{\footnotesize{1. P--prograde, R--retrograde}}\\
  \end{tabular}
      \end{table}

      The critical semi-major axis ratios were then regressed against the parameters discussed in the previous section using linear regression to extract semi-empirical relationships of the form
      {$$\frac{a_{\mathrm{io}}}{a_{2}} ,\frac{a_{\mathrm{oi}}}{a_{2}}=C+b_{1}\mu_{1}+b_{2}\mu_{2}+b_{3}e_{1}+b_{4}e_{2}+b_{5}i_{2}+b_{6}\Omega_{2}+b_{7}\omega_{2}\quad(1) $$}for the outer bounds of S1, S3 and P1 orbits, and the inner bounds of P2 orbits, respectively, where $C$ and $b_{i}$ are the regression constant and coefficients. In all the regressions of P-type and S-type orbits, the univariate relationships between the critical ratios and the independent variables were linear, with one exception -- the relationship between $a_{\mathrm{io}}/a_{2}$  and $\mu_{2}$ in S3 orbits.

      The results of the regressions are tabulated in Table \ref{table:2}. In this and subsequent tables of regression results, the various parameters are defined as follows: $\sigma$-standard deviation of critical ratios; $R^{2}$-coefficient of determination, $F$-$F$-statistic for overall significance, $SE$-standard error and $MAPE$-mean average percentage error, all being for the overall regression fit; $t$-$t$-statistic for individual coefficients and $N$-number of data points.

      The various regressions resulted in 24 regression constants and 192 coefficients. The semi-major axis ratio $a$ is included as an error check as the critical ratio scales with it and its coefficient should be zero.

\begin{sidewaystable*}
\centering
\caption{Summary of regression statistics for all triple orbital configurations, for regression equations $a_{\mathrm{io}}/{a_{2}},a_{\mathrm{oi}}/{a_{2}}=C+b_{1}\mu_{1}+b_{2}\mu_{2}+b_{3}e_{1}+b_{4}e_{2}+b_{5}i_{2}+b_{6}\Omega_{2}+b_{7}\omega_{2}$.}
\label{table:2}
\setlength{\tabcolsep}{1.5pt} % Default value: 6pt
\renewcommand{\arraystretch}{1.1} % Default value: 1
\begin{tabular}{lclcccccccccccccccrrccrrr} \hline
\setlength\tabcolsep{3pt}Orbit & Crit. & High${}^{1}$ & \multicolumn{2}{c}{Motions${}^{2}$} & \multicolumn{4}{c}{Crit. semi-major axis ratio} & \multicolumn{9}{c}{Regression coefficients} & \multicolumn{3}{c}{Integrations} & \multicolumn{4}{c}{ Model fit} \\
 type & ratio & incl. & Star 3 & Planet & Min & Mean & $\sigma$ & Max & $C$ & $a$ & $\mu_{1}$ & $\mu_{2}$ & $e_{1}$ & $e_{2}$ & $i_{2}$ & $\Omega_{2}$ & $\omega_{2}$ & No. & Bounds & Rate & $R^{2}$ & $F$ & SE & MAPE \\
 &  &  &  &  &  &  &  &  &  &  &  &  &  &  &  &  &  &  & found & (\%)  &  &  &  & (\%) \\ \hline
P1 & ${a_{\mathrm{io}}/{a_{2}}}$ &  & P  & P & 0.131 & 0.383 & 0.147 & 0.892 & 0.439 & 0.000 & -0.005 & -0.020 & -0.004 & -0.114 & 0.000 & -0.001 & -0.001 & 5\,413 & 1\,782 & 33  & 0.132 & 34    & 0.093 & 19 \\
 &  &  &  & R & 0.113 & 0.519 & 0.108 & 0.828 & 0.572 & 0.000 & 0.009 & -0.043 & \textit{-0.010} & -0.468 & 0.000 & 0.000 & 0.000 & 5\,343 & 976   & 18  & 0.670 & 241   & 0.062 & 10 \\ \cline{4-25}
 &  &  & R & P & 0.110 & 0.385 & 0.077 & 0.897 & 0.463 & 0.000 & 0.192 & -0.037 & \textit{-0.021} & -0.187 & -0.001 & 0.000 & 0.000 & 2\,122 & 569   & 27  & 0.547 & 285   & 0.358 & 9 \\
 &  &  &  & R & 0.112 & 0.537 & 0.132 & 0.857 & 0.656 & 0.000 & \textit{-0.056} & -0.022 & 0.005 & -0.510 & 0.000 & 0.000 & 0.000 & 2\,685 & 539   & 20  & 0.710 & 162   & 0.071 & 12 \\ \cline{3-25}
 &  & NK & P & P & 0.049 & 0.484 & 0.099 & 0.837 & 0.545 & 0.000 & -0.044 & -0.018 & 0.000 & 0.000 & -0.001 & 0.000 & 0.000 & 342   & 256   & 75  & 0.024 & 2     & 0.099 & 18 \\
 &  & K &  & P & 0.288 & 0.495 & 0.100 & 0.748 & 0.303 & 0.001 & -0.170 & 0.017 & 0.000 & 0.000 & 0.002 & 0.000 & 0.000 & 319   & 47    & 15  & 0.177 & 2     & 0.095 & 19 \\ \cline{3-25}
 &  & NK & R & P & 0.423 & 0.666 & 0.078 & 0.853 & 0.685 & -0.001 & 0.066 & 0.029 & 0.000 & 0.000 & -0.001 & 0.000 & 0.000 & 273   & 178   & 65  & 0.202 & 4     & 0.024 & 13 \\
 &  & K &  & P & 0.105 & 0.531 & 0.096 & 0.818 & 0.738 & 0.000 & -0.038 & -0.083 & 0.000 & 0.000 & -0.001 & 0.000 & 0.000 & 207   & 102   & 49  & 0.194 & 6     & 0.088 & 12 \\ \hline
 P2 & ${a_{\mathrm{oi}}/{a_{2}}}$ &  & P  & P & 1.253 & 2.936 & 1.449 & 5.197 & 2.376 & -0.003 & 0.053 & 0.044 & 0.090 & 1.996 & 0.000 & -0.001 & -0.002 & 5\,413 & 5\,133 & 95  & 0.829 & 3\,097 & 0.248 & 6 \\
 &  &  &  & R & 1.184 & 1.976 & 0.507 & 4.916 & 1.482 & -0.004 & 0.131 & -0.008 & \textit{-0.036} & 1.727 & 0.001 & 0.001 & 0.001 & 5\,343 & 5\,082 & 95  & 0.865 & 3\,910 & 0.198 & 6 \\ \cline{4-25}
 &  &  & R & P & 0.924 & 2.773 & 0.891 & 5.384 & 2.089 & -0.004 & 0.809 & 0.114 & 0.015 & 1.298 & -0.001 & 0.000 & 0.000 & 2\,122 & 1\,895 & 89  & 0.271 & 26    & 0.126 & 31 \\
 &  &  &  & R & 0.591 & 1.960 & 0.571 & 4.957 & 1.708 & -0.005 & 0.182 & 0.023 & 0.018 & 1.693 & -0.001 & 0.000 & 0.000 & 2\,685 & 2\,455 & 91  & 0.773 & 1\,039 & 0.269 & 8 \\ \cline{3-25}
 &  & NK & P & P & 1.755 & 1.941 & 0.051 & 2.242 & 1.920 & -0.002 & 0.048 & 0.016 & 0.000 & 0.000 & 0.002 & 0.000 & 0.000 & 342   & 341   & 100  & 0.269 & 30    & 0.044 & 4 \\
 &  & K &  & P & 1.759 & 1.933 & 0.176 & 2.532 & 2.345 & -0.004 & 0.066 & 0.043 & 0.000 & 0.000 & -0.003 & 0.000 & 0.000 & 319   & 319   & 100 & 0.589 & 117   & 0.115 & 3 \\ \cline{3-25}
 &  & NK & R & P & 1.095 & 1.335 & 0.127 & 1.820 & 1.288 & -0.002 & 0.085 & -0.004 & 0.000 & 0.000 & 0.007 & 0.000 & 0.000 & 273   & 273   & 100 & 0.555 & 83    & 0.085 & 4 \\
 &  & K &  & P & 1.363 & 1.753 & 0.195 & 2.510 & 1.356 & -0.003 & 0.033 & -0.016 & 0.000 & 0.000 & 0.008 & 0.000 & 0.000 & 207   & 204   & 99  & 0.574 & 64    & 0.129 & 5 \\ \hline
 S1 & ${a_{\mathrm{io}}/{a_{1}}}$ &  & P  & P & 0.015 & 0.180 & 0.049 & 0.760 & 0.390 & 0.000 & -0.746 & 0.005 & -0.398 & -0.006 & 0.000 & 0.000 & 0.000 & 1\,532 & 819   & 54  & 0.583 & 142   & 0.081 & 35 \\
 &  &  &  & R & 0.015 & 0.185 & 0.122 & 0.771 & 0.392 & 0.000 & -0.667 & 0.003 & -0.381 & 0.002 & 0.000 & 0.000 & 0.000 & 1\,480 & 785   & 53  & 0.567 & 131   & 0.081 & 37 \\ \cline{4-25}
 &  &  & R & P & 0.015 & 0.197 & 0.059 & 0.772 & 0.277 & 0.000 & -0.387 & -0.005 & -0.125 & -0.005 & 0.000 & 0.000 & 0.000 & 1\,586 & 957   & 60  & 0.126 & 17    & 0.139 & 67 \\
 &  &  &  & R & 0.028 & 0.196 & 0.157 & 0.772 & 0.423 & 0.000 & -0.451 & -0.006 & -0.056 & -0.048 & 0.001 & 0.000 & 0.000 & 1\,455 & 791   & 54  & 0.176 & 20    & 0.143 & 74 \\ \hline
 S3 & ${a_{\mathrm{io}}/{a_{2}}}$ &  & P  & P & 0.009 & 0.289 & 0.098 & 0.893 & 0.393 & 0.000 & -0.075 & 0.037 & -0.020 & -0.604 & 0.000 & 0.000 & 0.000 & 1 604 & 1\,599 & 100& 0.712 & 481   & 0.084 & 25 \\
 &  &  &  & R & 0.010 & 0.361 & 0.158 & 0.920 & 0.444 & 0.001 & 0.001 & 0.113 & -0.011 & -0.580 & 0.000 & 0.000 & 0.000 & 1\,562 & 1\,501 & 96  & 0.790 & 705   & 0.071 & 17 \\ \cline{4-25}
 &  &  & R & P & 0.007 & 0.287 & 0.096 & 0.853 & 0.391 & 0.002 & -0.071 & 0.033 & -0.001 & -0.597 & 0.000 & 0.000 & 0.000 & 1\,596 & 1\,591 & 100  & 0.727 & 522   & 0.077 & 23 \\
 &  &  &  & R & 0.009 & 0.360 & 0.156 & 0.908 & 0.425 & 0.000 & -0.002 & 0.114 & -0.002 & -0.580 & 0.000 & 0.000 & 0.000 & 1\,537 & 1\,489 & 97  & 0.790 & 688   & 0.071 & 17 \\ \hline
 \multicolumn{3}{l}{Average$^{3}$/total$^{4}$} &  &  & 0.473 & 0.944 & 0.237 & 1.789 & 0.921 & 0.000 & 0.183 & 0.036 & 0.050 & 0.433 & 0.001 & 0.000 & 0.000 & 45\,760 & 29\,683 & 65  & 0.495 & -   & -   & -   \\ \hline
\end{tabular}
\tablefoot{ \\
\tablefoottext{1}{NK--non-Kozai, K--Kozai} \\
\tablefoottext{2}{P--prograde, R--retrograde} \\
\tablefoottext{3}{Of absolute values for regression coefficients} \\
\tablefoottext{4}{Of integrations and bounds} \\
\tablefoottext{5}{The directional movements of the stability bounds for planetary orbits are generally as expected, except for retrograde inner planetary orbits, for which the bound increases instead of shrinking. This may be attributable to the small sample, since very few retrograde star/retrograde planet combinations were stable. Italics denote coefficients with signs that were not as expected. Ignoring coefficients smaller than 0.01, only four did not have the expected sign and these were all small $(<|0.06|)$}. \\
\tablefoottext{6}{The semi-major axis ratio $a$ is included as an error check as the critical ratio scales with it and its coefficient should be zero. Its average absolute value of $<0.0005$ therefore provides an indication of the intrinsic error of the methodology.} \\
}
\end{sidewaystable*}
      Examining the constants, for circular, coplanar outer stellar orbits, the inner and outer stability boundaries for prograde planetary orbits are found at around 0.4 times and 2.4 times the distance of the outer star respectively, and for retrograde planetary orbits, at around 0.6 times and 1.5 times respectively. The parameters $i_{2}$, $\Omega_{2}$ and $\omega_{2}$ have a negligible influence on the critical ratio in these low-inclination cases.

      For P1 orbits, the dominant influence on the critical semi-major axis ratio is only the outer star’s eccentricity, particularly for retrograde planetary orbits. For P2 orbits this effect is even stronger, for both planetary motions. In the case of a retrograde outer star these influences are largely unchanged for P1 orbits, but for P2 orbits the inner mass ratio becomes important for a prograde planet.

      The outer star dominates the regions of stability in a triple system, with its eccentricity having by far the largest influence. The configuration of the inner binary has little effect on either inner or outer orbits. This results from the fact that, as a consequence of the Mardling stability limit, the outer star is sufficiently far away that the inner binary effectively resembles a single point mass. These conclusions apply to both prograde and retrograde planetary orbits. However, the greater stability of retrograde planetary orbits results in inner and outer bounds that are closer to the outer star compared with the prograde case.

      The regression coefficient of the outer eccentricity $e_{2}$ is large -- for highly eccentric orbits of the outer star, the critical semi-major axis ratio can expand by over 80\% for prograde planetary orbits and more than double for retrograde orbits. The inner bound can shrink by a quarter for prograde planetary orbits and by over 80\% for retrograde orbits.

      The effect of the direction of motion of the outer star on the planetary stability bounds is shown in Table \ref{table:3}. The significance level is shown for the mean critical ratio, and differences in the regression constant are shown for comparison. Although the absolute differences in mean critical ratio are small, two are statistically significant.

      \begin{table}[htbp]
      \centering
      \caption{Differences in mean critical semi-major axis ratios and regression constants for P1 and P2 orbits, for retrograde relative to prograde stellar motion, with the same planetary motion.}
      \label{table:3}
      \begin{tabular}{*{5}{c}}
      \hline
      Orbit & Ratio & Planet & Difference in & Difference in \\
      type &  & motion & mean critical & regression \\
      &  &  &  ratio* & constant\\
      &  &  & (\%) & (\%) \\ \hline
      P1 & ${a_{\mathrm{io}}/{a_{2}}}$ & P & 1 & 6 \\
      & & R & 3* & 15 \\ \hline
      P2 & ${a_{\mathrm{oi}}/{a_{2}}}$ & P & -6* & -12 \\
      & & R & -1 & 15 \\ \hline
      \multicolumn{5}{p{2in}}{\footnotesize{* significant at the 5\% level}} \\
      \end{tabular}
      \end{table}

  The effect of the direction of motion of planets on their stability bounds is shown in Table \ref{table:4}. The absolute differences are all large, averaging 37\% for P1 orbits and 31\% for P2 orbits, with the signs being consistent with the greater stability of retrograde orbits.

  \begin{table}[htbp]
      \centering
      \caption{Differences in mean critical semi-major axis ratios and regression constants for P1 and P2 orbits, for retrograde relative to prograde planetary motion, with the same stellar motion.}
      \label{table:4}
      \begin{tabular}{*{5}{c}}
  \hline
  Orbit & Ratio & Star 3 & Difference in & Difference in \\
  type &  & motion & mean critical & regression \\
  &  &  &  ratio* & constant\\
  &  &  & (\%) & (\%) \\ \hline
  P1 & ${a_{\mathrm{io}}/{a_{2}}}$ & P & 36* & 31 \\
  & & R & 39* & 42 \\ \hline
  P2 & ${a_{\mathrm{oi}}/{a_{2}}}$ & P & -33* & -38 \\
  & & R & -29* & -18 \\ \hline
  \multicolumn{5}{p{2in}}{\footnotesize{* significant at the 5\% level}} \\
  \end{tabular}
    \end{table}

  \subsubsection{Stability bounds and the outer star’s eccentricity}
\begin{figure}[htbp]
     \centering
     \includegraphics[width=\hsize]{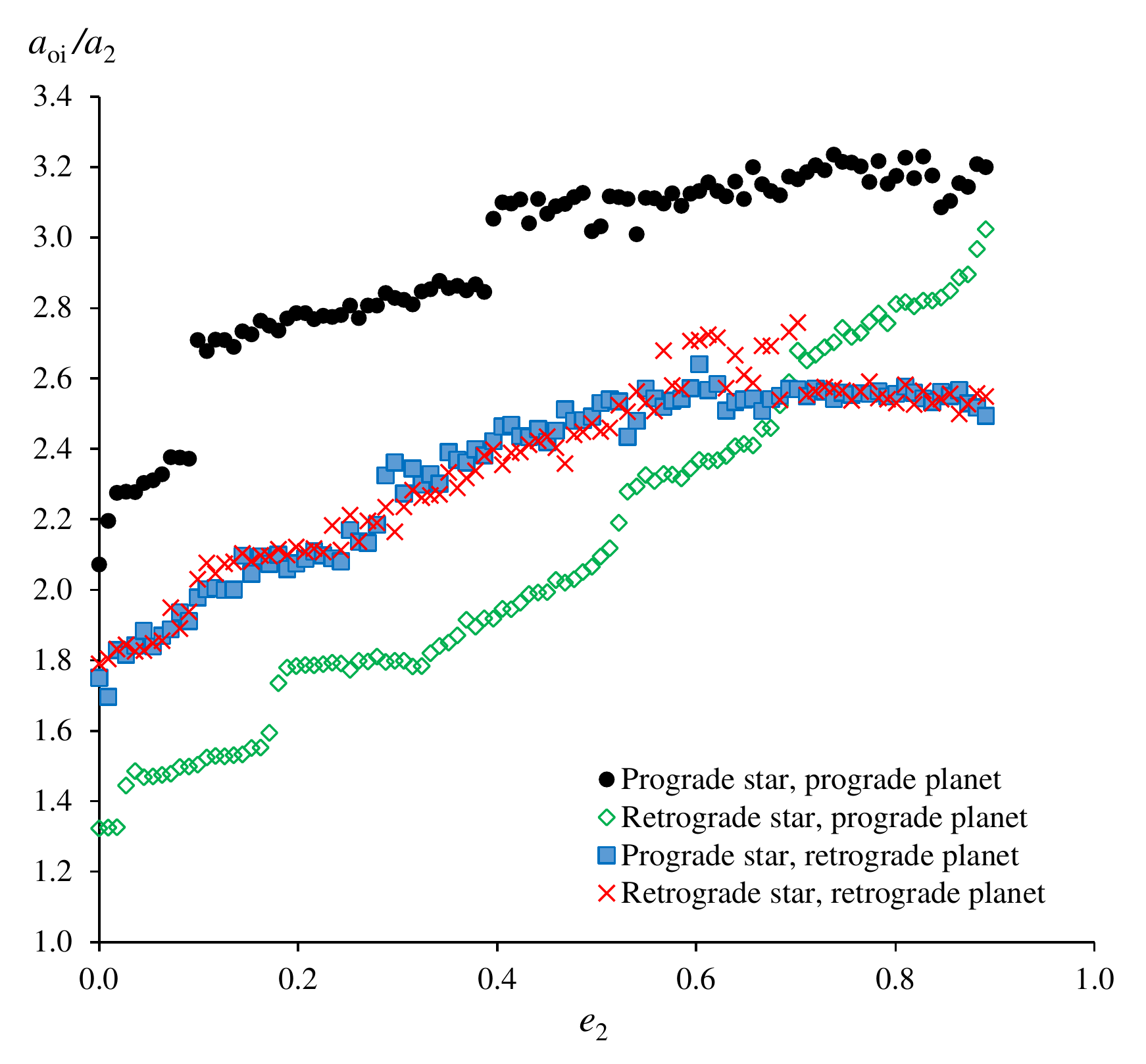}
        \caption{Outer stability bound for P2 orbits as a function of outer star eccentricity, for $a=100$ AU, $\mu_{1}=\mu_{2}=0.5, e_{1}=i_{1}=\Omega_{2}=\omega_2=0$ and $i_{2}=0^\circ$ and 180$^\circ$.}
  \label{Fig2}
     \end{figure}

  Since the eccentricity of the outer star is by far the most influential variable on both the inner and outer planetary stability bounds, a series of integrations was run to examine the relationship between these two variables.
  Typical relationships for the outer bounds are illustrated in Fig.\ref{Fig2}, for one stellar configuration.

     For the outer bound the critical ratio should be an increasing function of outer eccentricity $e_2$. This is true for all four cases shown, although for the two retrograde planet cases the critical ratio flattens out for $e_2\gtrsim0.6$. There are discontinuities in the critical ratio, resulting from resonances, in each case. This is most visible in the prograde star and prograde planet case, with gaps occurring at eccentricities of 0.10, where the critical ratio jumps from 2.37 to 2.70; and at 0.39, with the ratio undergoing a step change from 2.82 to 3.05. These instabilities are far less pronounced in the two retrograde planet cases.

     A retrograde outer body allows the outer bound for a prograde planet to move substantially closer to it, with a critical ratio at $e_2=0$ of $\sim$1.3, compared with $\sim$2.1 for a prograde outer star. However, this difference diminishes with increasing outer eccentricity, with both critical ratios converging toward
3.2 as this eccentricity approaches unity.

\begin{figure}[htbp]
     \centering
     \includegraphics[width=\hsize]{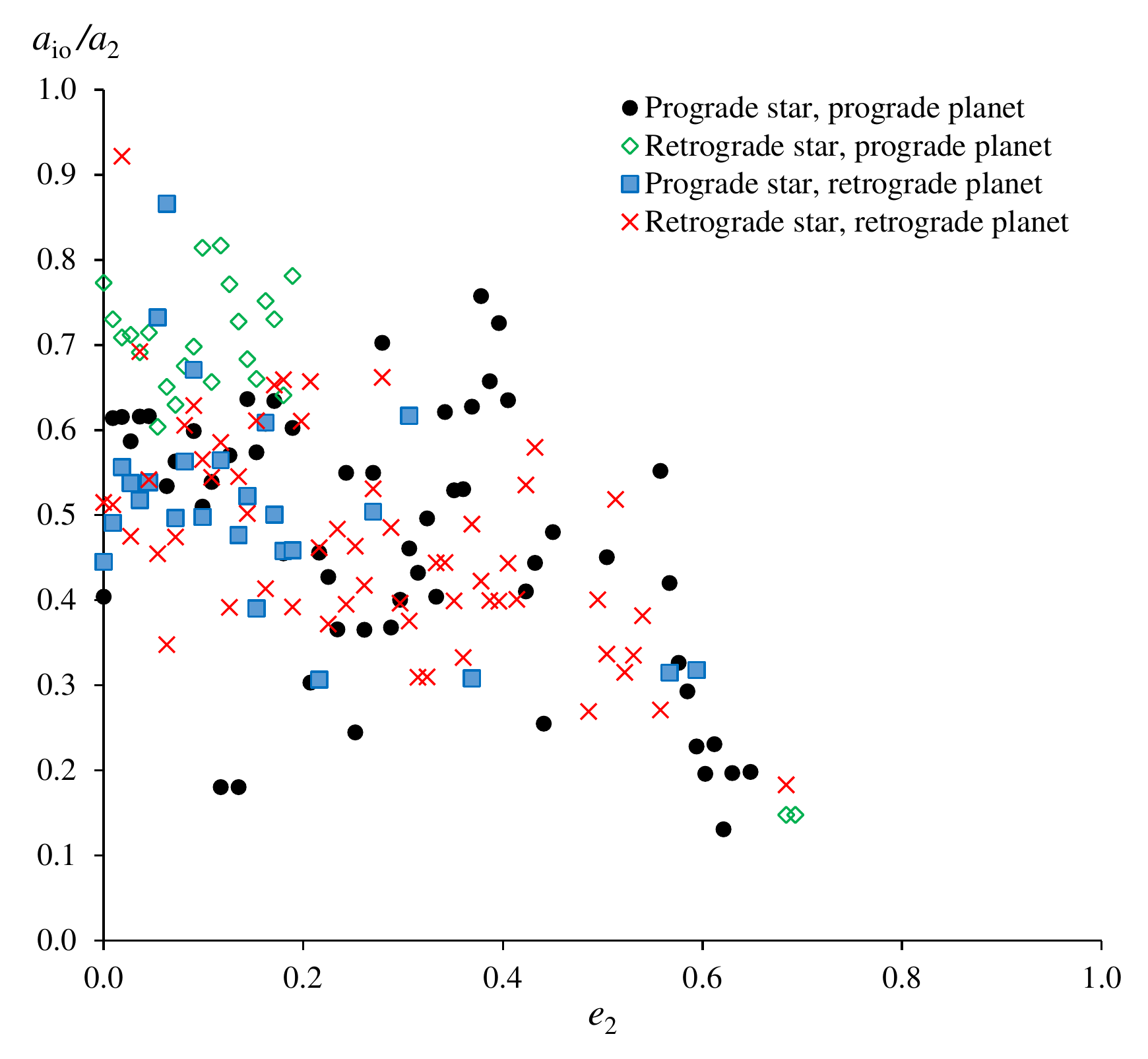}
        \caption{Inner stability bound for P1 orbits as a function of outer star eccentricity, for $a=100$ AU, $\mu_{1}=\mu_{2}=0.5, e_{1}=i_{1}=\Omega_{2}=\omega_2=0$ and $i_{2}=0^\circ$ and 180$^\circ$.}
  \label{Fig3}
     \end{figure}

  Typical relationships for the inner bounds are illustrated in Fig. \ref{Fig3}, for the same stellar configuration. For the inner bound the critical ratio is expected to be a decreasing function of outer eccentricity. This is the general trend in each case, albeit with high scatter, which reflects that the inner orbit bounds are always more diffuse than the outer orbit bounds. This is a result of the greater sparsity of surviving test particles compared with outer orbits, since inner orbits tend to be more chaotic. No stable inner orbits are found for outer eccentricities above 0.7.

  \subsubsection{Comparison with observed P1 and P2 orbits}
  The only P1 orbit found in a triple system is in HW Virginis, with a semi-major axis ratio of 0.37 \citep{beue2012}. No circumtriple P2 orbits have been discovered to date.
  The planet with the smallest P1 orbit in a binary, in terms of absolute semi-major axis, is Kepler 47b \citep{oros2012}, with a semi-major axis ratio (here relative to the binary) of ${a_{\mathrm{io}}/{a_{1}}}=3.54$. The smallest semi-major axis ratio of a planet’s orbit is for Kepler 16b, with ${a_{\mathrm{io}}/{a_{1}}}=3.14$ \citep{doyl2011}. Both lie well outside the smallest mean critical semi-major axis ratios of $\sim$0.1 found in the simulations, so the possibility exists of finding planets in even smaller orbits.

  \subsubsection{P1 and P2 orbits in triples compared with binaries}
  To highlight how the P1 and P2 planetary stability bounds in a triple differ from those in a binary, the integrations were repeated after reducing the triple to a binary by condensing the inner binary into a single body. The results of this approximation for the mean critical semi-major axis ratios for the prograde stellar case are compared in Table \ref{table:5}.

  \begin{table}[htbp]
  \centering
      \caption{Difference between mean critical semi-major axis ratios in triples and binaries, for P1 and P2 orbits.}
      \label{table:5}
      \setlength{\tabcolsep}{5.5pt} % Default value: 6pt
      \begin{tabular}{l|ccc|ccc}
  \hline
  Planetary & \multicolumn{3}{c|}{P1: ${a_{\mathrm{io}}/{a_{2}}}$} & \multicolumn{3}{c}{P2: ${a_{\mathrm{oi}}/{a_{2}}}$} \\
  orbit motion & Triple & Binary & $\Delta(\%)$ & Triple & Binary & $\Delta(\%)$ \\\hline
  Prograde & 0.383 & 0.364 & -5* & 2.936 & 2.703 & -8* \\
  Retrograde & 0.519 & 0.471 & -9* & 1.976 & 1.691 & -14* \\ \hline
  $\Delta$(\%) & 35 & 29 & - & -33 & -37 & - \\ \hline
  \multicolumn{7}{p{2in}}{\footnotesize{* significant at the 5\% level}} \\
  \end{tabular}
    \end{table}

  Both the inner and outer mean critical semi-major axis ratios move toward the central star, for both prograde and retrograde planetary orbits. While these movements are statistically significant, they are nevertheless small, confirming that the influence of the outer star is dominant.
  The regression coefficients are compared in Table \ref{table:6}.

      \begin{table}[htbp]
      \centering
      \caption{Regression coefficients and model fits for P1 and P2 orbits, triples compared with binaries, for regression equations $a_{\mathrm{io}}/{a_{2}},a_{\mathrm{oi}}/{a_{2}}=C+b_{1}\mu_{1}+b_{2}\mu_{2}+b_{3}e_{1}+b_{4}e_{2}+b_{5}i_{2}+b_{6}\Omega_{2}+b_{7}\omega_{2}$.}
      \label{table:6}
\begin{tabular}{lrrrrr} \hline
Parameter & \multicolumn{2}{c}{Triple} & \multicolumn{2}{c}{Binary} & $\Delta$Coeff. \\ \cline{2-5}
  & Coeff. & $t$ & Coeff. & $t$ & (\%) \\ \hline
\multicolumn{6}{c}{Inner orbits, prograde planet} \\ \hline
$C$ & 0.439 & 47.8 & 0.195 & 19.1 & -55 \\
$a$ & 0.000 & 0.4 & 0.000 & 47.9 & 11 \\
$\mu_{1}$ & -0.005 & -0.4  & - & - & - \\
$\mu_{2}$ & -0.020 & -7.1  & -0.006 & -1.8  & -68 \\
$e_{1}$ & -0.004 & -0.5  & -  & -  & - \\
$e_{2}$ & -0.114 & -8.9  & -0.435 & -39.4 & 283 \\
$i_{2}$ & 0.000 & 2.2 & -  & -  & - \\
$\Omega_{2}$ & -0.001 & -6.1  & -  & -  & - \\
$\omega_{2}$ & -0.001 & -6.3  & - & - & - \\ \hline
$R^{2}$ & 0.132 &  & 0.563 &  & \\
MAPE (\%) & 19 &  & 11 &  & \\ \hline
\end{tabular}
\begin{tabular}{lrrrrr}
\multicolumn{6}{c}{Outer orbits, prograde planet} \\ \hline
$C$ & 2.376 & 195.6 & 2.719 & 84.2 & 14 \\
$a$ & -0.003 & -23.8 & 0.000 & -25.8 & -98 \\
$\mu_{1}$ & 0.053 & 2.9  & - & - & - \\
$\mu_{2}$ & 0.044 & 11.5 & -0.017 & -1.4  & -138 \\
$e_{1}$ & 0.090 & 6.8 & - & - & - \\
$e_{2}$ & 1.996 & 151.1 & 0.884 & 25.4 & -56 \\
$i_{2}$ & 0.000 & 3.1 & -  & -  & - \\
$\Omega_{2}$ & -0.002 & -15.9 & - & - & - \\
$\omega_{2}$ & -0.002 & -13.4 & - & - & - \\ \hline
$R^{2}$ & 0.829 &  & 0.893 &  & \\
MAPE (\%) & 6 &  & 3 &  \\ \hline
\end{tabular}
\begin{tabular}{lrrrrr}
\multicolumn{6}{c}{Inner orbits, retrograde planet} \\ \hline
$C$ & 0.572 & 76.0  & 0.211 & 14.3  & -63 \\
$a$ & 0.000 & 4.8   & 0.000 & 31.6  & -90 \\
$\mu_{1}$ & 0.009 & 0.8 & - & - & - \\
$\mu_{2}$ & -0.043 & -19.2 & -0.026 & -4.9 & -39 \\
$e_{1}$ & -0.010 & -1.4 & - & - & - \\
$e_{2}$ & -0.468 & -31.3 & -0.567 & -34.9 & 21 \\
$i_{2}$ & 0.000 & -2.8  & - & - & - \\
$\Omega_{2}$ & 0.000 & 1.0 & - & - & - \\
$\omega_{2}$ & 0.000 & 1.3 & - & - & - \\ \hline
$R^{2}$ & 0.670 &  & 0.444 &  &  \\
MAPE (\%) & 10 &  & 13 &  & \\ \hline
\end{tabular}
\begin{tabular}{lrrrrr}
\multicolumn{6}{c}{Outer orbits, retrograde planet} \\ \hline
$C$ & 1.482 & 153.1 & 1.235 & 141.9 & -17 \\
$a$ & -0.004 & -40.4 & 0.000 & -45.7 & -99 \\
$\mu_{1}$ & 0.131 & 9.1 & - & - & - \\
$\mu_{2}$ & -0.008 & -2.8 & 0.023 & 7.3 & -375 \\
$e_{1}$ & -0.036 & -3.5 & - & - & - \\
$e_{2}$ & 1.727 & 166.6 & 1.681 & 175.0 & -3 \\
$i_{2}$ & 0.001 & 19.6  & - & - & - \\
$\Omega_{2}$ & 0.001 & 8.5 & - & - & - \\
$\omega_{2}$ & 0.001 & 5.1 & - & - & - \\ \hline
$R^{2}$ & 0.865 &  & 0.921 &   &  \\
MAPE (\%) & 6 &  & 4  &  &  \\ \hline
\end{tabular}
     \end{table}

  In all cases, aside from the constant the only variable of major influence is the eccentricity of the outer star, $e_{2}$. Generally, the effect of the outer star’s eccentricity is far larger on the outer stability bounds than on the inner stability bounds. For inner orbits its influence is larger in binaries than in triples, for both prograde and retrograde planetary orbits. For outer orbits, its effect in triples and binaries is essentially the same for retrograde orbits, but for prograde orbits it has a larger influence in triples than in binaries. Compared with binaries, the influence of the outer star’s eccentricity is significantly smaller for inner prograde orbits and materially larger for outer prograde orbits; there are no similar differences for retrograde orbits. Overall, the relatively small differences between triples and binaries result from the Mardling stability limit for triples, which precludes them from becoming too compact.

  \subsection{Orbit types P1 and P2 in highly inclined triple systems}
  \subsubsection{Vertical characteristics of the stability region}
\begin{figure}[htbp]
     \centering
     \includegraphics[width=\hsize]{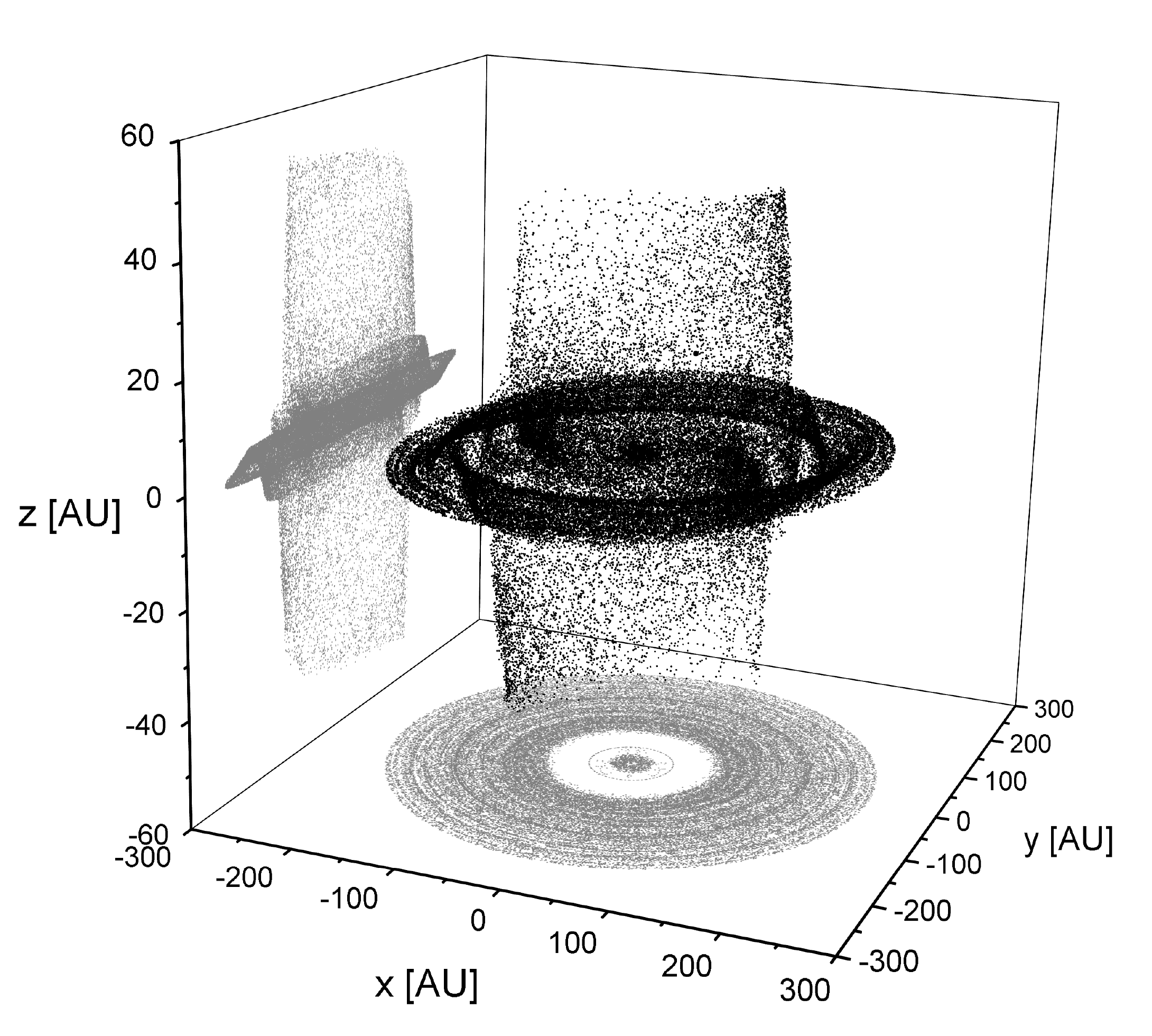}
        \caption{Position in space of surviving test particles after 100 kyr. P1 and P2 prograde orbits, with $a=40$ AU, $e_{1}=e_{2}=0$, $i_{1}=0^\circ$, $i_{2}=65^\circ$, $\mu_{1}=0.5$, $\mu_{2}=2.0$.}
  \label{Fig4}
     \end{figure}

  Here we examine the stable planetary region for large inclinations of the outer star relative to the invariable plane of the triple system. Once the orbit of the outer star is no longer coplanar with the orbit of the inner binary, in other words, the mutual inclination of the two orbits is no longer zero, the stable planetary region will also no longer be coplanar but will extend vertically, with its shape being sculpted primarily by the outer star.
  An example is illustrated in Fig. \ref{Fig4}, where the $x$--$y$ plane is aligned with the invariable plane. The outer limit of the test particle cloud is of no relevance here, since this is simply where it has been truncated.

 Examining the paths of individual test particles over time yields the following observations. Firstly, their orbits librate, with the argument of periapsis $\omega$ oscillating. Secondly, the semi-major axis of their orbits, $a$, remains effectively constant. Thirdly, their orbits remain circular, with their eccentricity staying very close to zero. Fourthly, their orbits vary in inclination. In an integration they are initially coplanar with the invariable plane and then become more inclined over time, eventually oscillating around a final non-zero inclination with a period much longer than those of the other orbital elements. The typical annular vertical “chimney” shown in Fig \ref{Fig4} therefore consists of test particles in circular orbits of constant semi-major axis but varying inclination. Fifthly, some test particles develop Kozai resonance and some of these eventually undergo orbit flipping into retrograde motion.

\begin{figure}[htbp]
     \centering
     \includegraphics[width=\hsize]{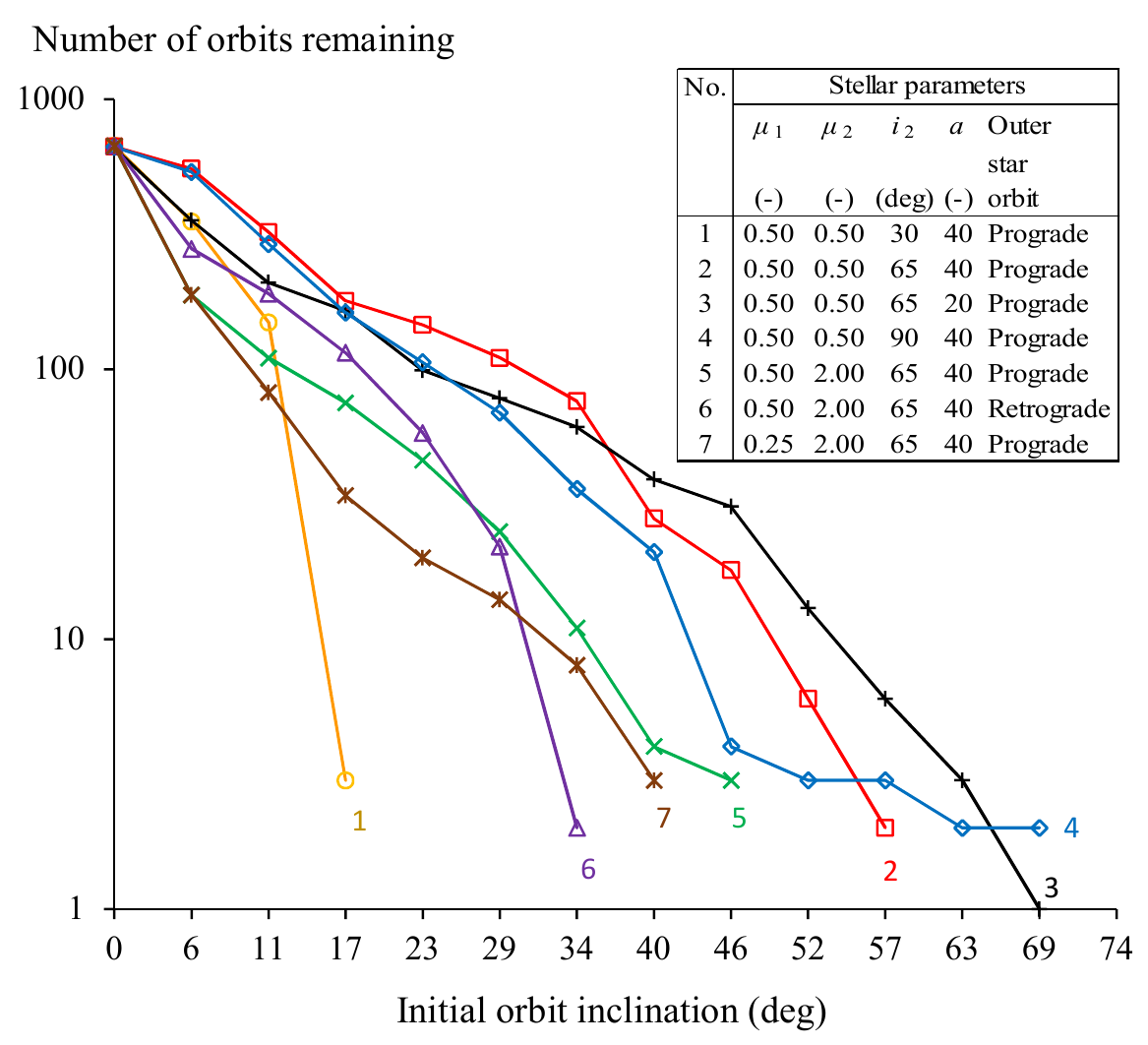}
        \caption{Number of stable P1 and P2 prograde orbits remaining after 1 Myr as a function of initial planetary inclination, for various triple configurations.}
  \label{Fig5}
     \end{figure}

  There is a limit to the maximum inclination of planetary orbits. The number of stable orbital bounds falls exponentially with increasing inclination, as shown in Fig. \ref{Fig5}, where the lines end when there are no more stable orbital bounds.

  \subsubsection{Planetary stability bounds and the outer star’s inclination}
    Both prograde and retrograde stellar orbits were used, and planetary motions were prograde. In practical terms, we know that for real bodies, as opposed to test particles, the Kozai critical inclination is larger than the theoretical 39$^\circ$, for example \citet{gris2017}. Also, if the ratio of the initial angular momentum of the outer orbit to that of the inner orbit is $\gtrsim$4 then significant Kozai resonance usually occurs, for example \citet{beus2012}.

\begin{figure}[htbp]
     \centering
     \includegraphics[width=\hsize]{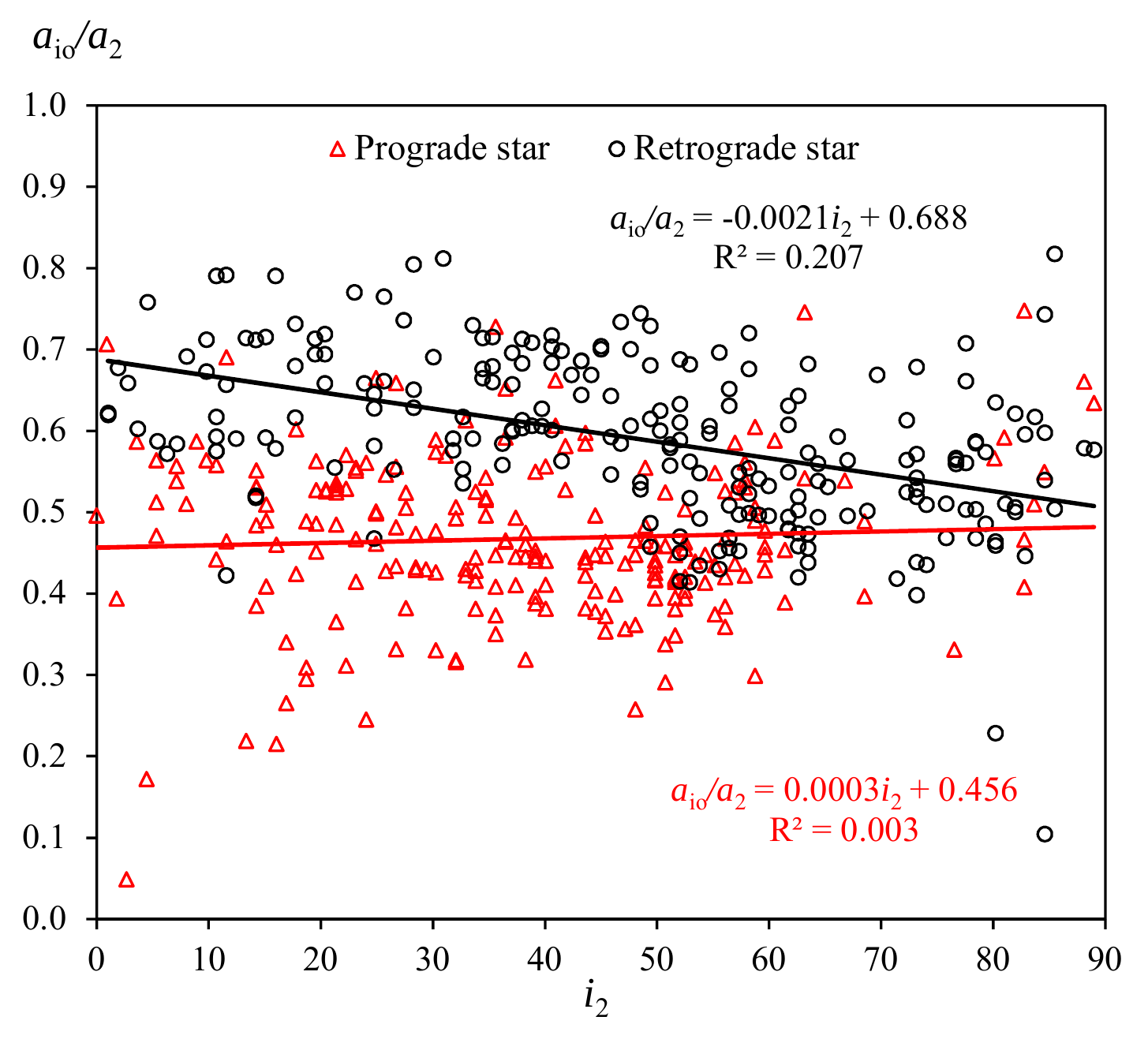}
     \includegraphics[width=\hsize]{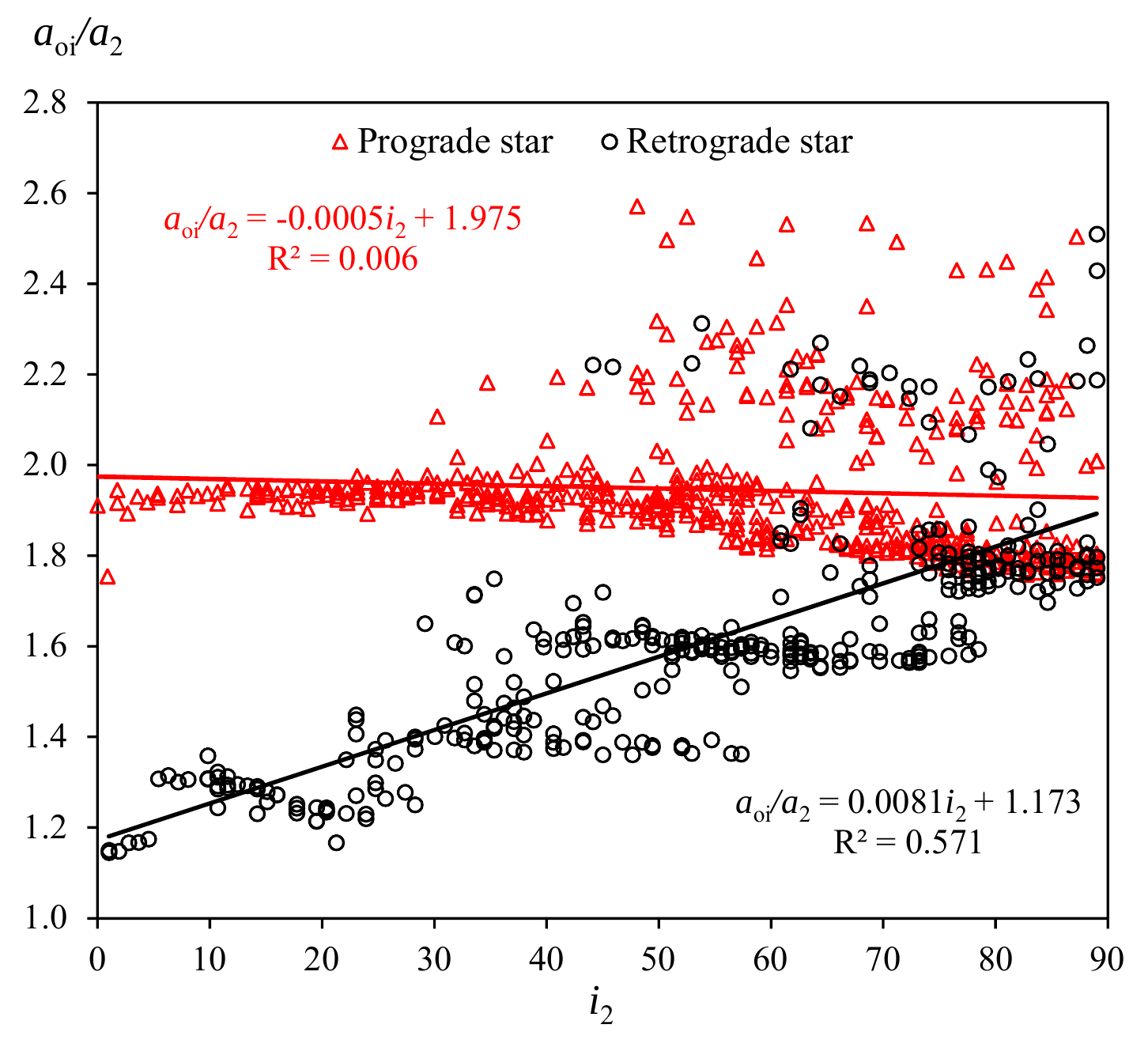}
        \caption{Critical semi-major axis ratios versus outer star inclination: panel a) inner stability bound; panel b) outer stability bound.}
  \label{Fig6}
     \end{figure}

 Since we used a relatively massive outer star, the integrations were split, not by the theoretical Kozai critical inclination, but into what were experimentally determined to actually be non-Kozai and Kozai regimes, that is, those of relatively low inclination whose nodes circulate and those of high inclination where the nodes librate about 90$^\circ$ or 270$^\circ$. For the integrations the sample was first split into prograde and retrograde stellar cases. Each case was then split into subsets where Kozai resonance did or did not occur. Figure \ref{Fig6} shows the critical semi-major axis ratios for the inner and outer planetary orbits against the “absolute” inclination $i_{2}$ (i.e., $i_{2}$ for prograde orbits and $180^\circ-i_{2}$ for retrograde orbits), for both Kozai and non-Kozai regimes. As usual, there is more scatter for inner orbits than outer orbits.

     For inner orbits, the critical semi-major axis ratio has almost no dependence on the outer star's inclination for prograde stellar orbits, and only a weak one for retrograde orbits. Retrograde stellar orbits allow the stable planetary region to extend further out (with a regression constant of 0.69 versus 0.46), but as inclination approaches $90^\circ$, this effect decreases. (The regression lines for inner and outer orbits should ideally coincide at an inclination of $90^\circ$.)

  For outer orbits, the critical ratio also has minimal dependence on inclination for prograde stellar orbits. However, it can be seen in the bottom panel of Fig. \ref{Fig6} that this low average dependence is made up of two parts. For low inclinations the critical ratio is relatively constant at 1.9--2.0, but as inclinations increase the critical ratio begins to decline, in other words, the orbital stability bound begins to move inwards. This makes intuitive sense because once Kozai resonance begins, as the outer star’s inclination increases, the eccentricity of the inner binary decreases, which allows this to occur. This divergence begins at around $45^\circ$ and is probably a manifestation of the start of Kozai resonance. This suggests that for real bodies which may also be relatively large, Kozai resonance is more likely to begin when $i_{2}\gtrsim45^\circ$ than for $i_{2}>39^\circ$.

  For retrograde stellar orbits dependence on $i_{2}$ is much stronger. In this case planetary orbits can again approach much closer to the outer star (with a regression constant of 1.2 versus 1.9), but this increased stability declines as inclination rises to $90^\circ$. Also of interest are the regions of stability interspersed with gaps of instability.

      \begin{table}[htbp]
      \centering
      \caption{Non-Kozai and Kozai cases -- mean critical semi-major axis ratios for prograde and retrograde stellar orbits, for P1 and P2 orbits.}
      \label{table:7}
      \setlength{\tabcolsep}{4pt} % Default value: 6pt
      \begin{tabular}{llrccccc} \hline
Orbit & Regime & $N$ & Mean & $\Delta(\%)$ & $\sigma$ & Min & Max\\
type &   &   & critical &   &   &   \\
 &   &   &  ratio &   &   &   \\  \hline
\multicolumn{8}{c}{Prograde outer star} \\ \hline
Inner & NK & 256 & 0.484 &  & 0.099 & 0.049 & 0.837 \\
${a_{\mathrm{io}}/{a_{2}}}$ & K & 47 & 0.495 & 2.2 & 0.100 & 0.288 & 0.748 \\ \hline
Outer & NK & 341   & 1.941 &    & 0.051 & 1.755 & 2.242 \\
${a_{\mathrm{oi}}/{a_{2}}}$ & K & 317   & 1.933 & -0.4  &  0.176 & 1.759 & 2.532 \\ \hline
\multicolumn{8}{c}{Retrograde outer star} \\ \hline
Inner & NK & 177   & 0.666 &       & 0.078 & 0.423 & 0.853 \\
${a_{\mathrm{io}}/{a_{2}}}$ & K & 101   & 0.531 & -20.4*     & 0.096 & 0.105 & 0.818 \\ \hline
Outer & NK & 272   & 1.335 &    & 0.126 & 1.095 & 1.820 \\
${a_{\mathrm{oi}}/{a_{2}}}$ & K & 204   & 1.753 & 31.3*     & 0.195 & 1.363 & 2.510 \\ \hline
\multicolumn{5}{p{2in}}{\footnotesize{NK--non-Kozai, K--Kozai}} \\
\multicolumn{5}{p{2in}}{\footnotesize{* significant at the 5\% level}} \\
\end{tabular} \\
   \end{table}

  The mean semi-major axis ratios for the non-Kozai and Kozai cases are shown in Table \ref{table:7}. Generally, with a retrograde outer star the inner orbits move outwards toward this star and the outer orbits move inwards, reflecting the greater stability of retrograde orbits. For a prograde outer star the existence of Kozai resonance makes no difference to the critical ratios of either the inner or outer bounds, which remain virtually identical. For a retrograde outer star, however, the bounds are very different. When Kozai resonance occurs the inner bound contracts inwards, while the outer bound moves outwards, and by a larger proportional amount. For both inner and outer orbits these movements are in a direction opposite to that usually induced by a retrograde stellar orbit. This shows, unsurprisingly, that Kozai resonance increases planetary instability.

  At the 5\% significance level, the critical semi-major axis ratios for planetary orbits under non-Kozai and Kozai regimes do not differ for a prograde outer body, but for the retrograde case they are substantially different, by an absolute 20\%--30\%. The number of stable inner orbits found under Kozai resonance with a prograde outer star is small.

  The regression data for these four cases is shown in Table \ref{table:2}. None of the orbital parameters have any significant influence on the critical semi-major axis ratio, with the possible exception of $\mu_{1}$ in the prograde star/Kozai case. This stability bound is therefore effectively represented by the constant. The data for the constant in the four cases is shown in Table \ref{table:8}.

      \begin{table}[htbp]
      \centering
      \caption{Regression constants for non-Kozai and Kozai regimes, for P1 and P2 orbits.}
      \label{table:8}
       \setlength{\tabcolsep}{2pt} % Default value: 6pt
\begin{tabular}{llrcccccc} \hline
Orbit & Regime & $N$ & Constant & $\Delta$ & Std. & $t$ & \multicolumn{2}{c}{95\% bounds} \\
type &   &   &   & (\%) &  Err. &   & Lower & Upper \\ \hline
\multicolumn{9}{c}{Prograde outer star} \\ \hline
Inner & NK & 256 & 0.545 &  & 0.032 & 17.2 & 0.483 & 0.607 \\
${a_{\mathrm{io}}/{a_{2}}}$ & K & 47 & 0.303 & -44 & 0.130 & 2.3 & 0.041 & 0.566 \\ \hline
Outer & NK & 341   & 1.92 &   & 0.01 & 207.6 & 1.90 & 1.94 \\
${a_{\mathrm{oi}}/{a_{2}}}$ & K & 317 & 2.35 & 22  &  0.05 & 47.7 & 2.25 & 2.44 \\ \hline
\multicolumn{9}{c}{Retrograde outer star} \\ \hline
Inner & NK & 177   & 0.685 &   & 0.024 & 28.0 & 0.637 & 0.733 \\
${a_{\mathrm{io}}/{a_{2}}}$ & K & 101 & 0.738 & 8 & 0.086 & 8.6 & 0.567 & 0.908 \\ \hline
Outer & NK & 272   & 1.29 &   & 0.02 & 67.4 & 1.25 & 1.33 \\
${a_{\mathrm{oi}}/{a_{2}}}$ & K & 204 & 1.36 & 5 & 0.07 & 18.3 & 1.21 & 1.50 \\ \hline
\multicolumn{9}{p{2in}}{\footnotesize{NK--non-Kozai, K--Kozai}} \\
\end{tabular} \\
   \end{table}

  For a prograde outer star, Kozai resonance results in the constant being 44\% smaller for inner orbits and 22\% larger for outer orbits. A retrograde outer body does not result in a significant change in either constant.

  \subsection{Orbit type S1}
  \subsubsection{Results}
  The S1 and S2 orbits around stars 1 and  2 respectively are interchangeable, so only one case is examined, denoted S1. The mean critical semi-major axis ratios for S1 orbits, for the four possible combinations of orbital motion, are shown in Table \ref{table:9}.

  \begin{table}[htbp]
  \centering
      \caption{Mean critical semi-major axis ratios for all combinations of orbital motions in S1 orbits.}
       \label{table:9}
       \setlength{\tabcolsep}{4pt} % Default value: 6pt
  \begin{tabular}{*{8}{c}}
  \hline
  Orbit & Critical & \multicolumn{2}{c}{Motions${}^{1}$} & \multicolumn{4}{c}{Mean
  critical ratio} \\
  type & ratio & Star 2 & Planet & Min & {Mean} & $\sigma$ & Max \\ \hline
  S1 & ${a_{\mathrm{io}}/{a_{1}}}$ & P & P & 0.015 & {0.180} & 0.049 & 0.760 \\
   & \textit{} & ~ & R & 0.015 & {0.185} & 0.122 & 0.771 \\ \cline{3-8}
   & \textit{} & R & P & 0.015 & {0.197} & 0.059 & 0.772 \\
   & \textit{} & ~ & R & 0.028 & {0.196} & 0.157 & 0.772 \\ \hline
  \multicolumn{8}{p{2in}}{\footnotesize{1. P--prograde, R--retrograde}}\\
  \end{tabular}
  \end{table}

  For S1 planetary orbits in triples, the mean critical semi-major axis ratio is in the range 0.180--0.197. S1 planetary orbits on average extend $\sim$8\% further out when the outer star is in a retrograde orbit. For prograde outer stellar orbits, retrograde planetary orbits extend slightly (3\%) further out while for retrograde stellar orbits they are virtually the same. The results of the corresponding regressions are shown in Table \ref{table:2}. The key determinants of the critical semi-major axis ratio for planetary S1 orbits are only the inner binary’s mass ratio and eccentricity, and for a retrograde outer star the influence of eccentricity is much weaker.

  \subsubsection{Comparison with observed S1 orbits}
  The two S1 orbits in triple systems found to date, HD 126614 A b and HD 2638 b, have semi-major axis ratios of 0.0649 and 0.00172 respectively. There is some uncertainty whether a third, Fomalhaut b, is a planet or a dust cloud or disk.

  Among the closest S1 orbits found in a binary are 0.7 AU for OGLE-2013-BLG-0341L b, with a semi-major axis ratio ${a_{\mathrm{io}}}/{a_{\mathrm{1}}}$ of 0.058--0.041 \citep{goul2014}; 1.09 AU for HD 59686 b, with ${a_{\mathrm{io}}}/{a_{\mathrm{1}}}$ $\sim$0.080 \citep{trif2018} and 0.382 AU for KOI-1257 b, giving ${a_{\mathrm{io}}}/{a_{\mathrm{1}}}$ $\sim$0.072 \citep{sant2014}. Although some of our integrations showed stable bounds with semi-major axis ratios as low as 0.015, they also extended to almost 0.8, with the greatest concentration in the region 0.03--0.25, so there appears to be significant scope for planets to be found further out from their host stars.

  \subsubsection{S1 orbits in triples compared with binaries}
  The triple was reduced to a binary by setting the outer star’s mass to a negligible value and the integrations were repeated. The results of this approximation for the mean critical semi-major axis ratios for the prograde stellar case are compared in Table \ref{table:13}.

  \begin{table}[htbp]
  \centering
      \caption{Difference between mean critical semi-major axis ratios in triples and binaries, for S1 orbits.}
      \label{table:13}
      \begin{tabular}{l|ccc}
  \hline
  Planetary & \multicolumn{3}{c}{S1: ${a_{\mathrm{io}}/{a_{1}}}$} \\
  orbit motion & Triple & Binary & $\Delta$(\%) \\ \hline
  Prograde & 0.180 & 0.257 & 43* \\
  Retrograde & 0.185 & 0.289 & 57* \\ \hline
  $\Delta$(\%) & 2 & 12 & - \\ \hline
  \multicolumn{4}{p{2in}}{\footnotesize{* significant at the 5\% level}} \\
  \end{tabular}
    \end{table}

  The outer star of a triple has a significant constraining influence on S1-type orbits compared with the binary case. In triples the added influence of the outer star makes S1 orbits move inwards, with the mean critical semi-major axis ratio reducing by over 30\%, from $\sim$0.270 to $\sim$0.183, while the difference in critical ratio between prograde and retrograde planetary orbits shrinks from 12\% to 2\%. The regressions for the binary case are compared with those for the triple case in Table \ref{table:14}.

      \begin{table}[htbp]
      \centering
      \caption{Regression coefficients and model fits for S1 orbits, triples compared with binaries, for regression equations $a_{\mathrm{io}}/{a_{2}},a_{\mathrm{oi}}/{a_{2}}=C+b_{1}\mu_{1}+b_{2}\mu_{2}+b_{3}e_{1}+b_{4}e_{2}+b_{5}i_{2}+b_{6}\Omega_{2}+b_{7}\omega_{2}$.}
      \label{table:14}
      \begin{tabular}{lcrcrr} \hline
Parameter & \multicolumn{2}{c}{Triple} & \multicolumn{2}{c}{Binary} & $\Delta$Coeff. \\ \cline{2-5}
  & Coeff. & $t$ & Coeff. & $t$ & (\%) \\ \hline
\multicolumn{6}{c}{S1 orbits, prograde planet} \\ \hline
$C$ & 0.390  & 28.4  & 0.360  & 54.4  & -8 \\
$a$ & 0.000  & 2.4   & 0.000  & 1.3   & -77 \\
$\mu_{1}$ & -0.746 & -32.4 & -0.497 & -50.7 & -33 \\
$\mu_{2}$ & 0.005  & 1.1   & 0.000  & 0.0   & - \\
$e_{1}$ & -0.398 & -19.7 & -0.396 & -58.8 & 0 \\
$e_{2}$ & -0.006 & -0.5  & -0.011 & -1.6  & 67 \\
$i_{2}$ & 0.000  & 1.6   & 0.000  & -0.3  & - \\
$\Omega_{2}$ & 0.000  & 1.4   & 0.000  & -0.9  & - \\
$\omega_{2}$ & 0.000  & 0.9   & 0.000  & 0.8   & - \\ \hline
$R^{2}$ & 0.583 &       & 0.680 &       &  \\
MAPE (\%) & 34.9  &       & 40.6  &       &  \\ \hline
\end{tabular}
\begin{tabular}{lcrcrr}
\multicolumn{6}{c}{S1 orbits, retrograde planet} \\ \hline
$C$ & 0.392  & 27.1  & 0.378  & 44.5  & -4 \\
$a$ & 0.000  & 2.8   & 0.000  & -1.7  & -136 \\
$\mu_{1}$ & -0.667 & -29.8 & -0.491 & -39.6 & -26 \\
$\mu_{2}$ & 0.003  & 0.6   & 0.000  & 0.0   & - \\
$e_{1}$ &  -0.381 & -18.9 & -0.411 & -47.9 & 7 \\
$e_{2}$ & 0.002  & 0.2   & 0.000  & 0.1   & -77 \\
$i_{2}$ & 0.000  & -0.2  & 0.000  & 1.9   & - \\
$\Omega_{2}$ & 0.000  & -0.4  & 0.000  & -0.1  & - \\
$\omega_{2}$ & 0.000  & 0.0   & 0.000  & -0.2  & - \\ \hline
$R^{2}$ & 0.567 &       & 0.641 &       &  \\
MAPE (\%) & 36.7  &       & 41.8  &       &  \\ \hline
\end{tabular}
   \end{table}

  The results for prograde and retrograde planetary orbits in triples and binaries are qualitatively similar, in that the only variables of influence are the inner mass ratio $\mu_{1}$ and eccentricity $e_{1}$. This effect is not attributable to Kozai, as the mutual inclination has no influence. However, the effect of the mass ratio is significantly larger in triples than in binaries, while that of eccentricity (and the constant) are effectively the same.

  \subsection{Orbit type S3}
  \subsubsection{Results}
  The mean critical semi-major axis ratios for S3 orbits, for the four possible combinations of orbital motion, are shown in Table \ref{table:15}.

  \begin{table}[htbp]
  \centering
      \caption{Mean critical semi-major axis ratios for all combinations of orbital motions in S3 orbits.}
      \label{table:15}
      \setlength{\tabcolsep}{4pt} % Default value: 6pt
      \begin{tabular}{*{8}{c}}
  \hline
  Orbit & Critical & \multicolumn{2}{c}{Motions${}^{1}$} & \multicolumn{4}{c}{Mean
  critical ratio} \\
  type & ratio & Star 3 & Planet & Min & Mean & $\sigma$ & Max \\ \hline
  S3 & ${a_{\mathrm{io}}/{a_{2}}}$ & P & P &  0.009 & {0.289} & 0.098 & 0.893 \\
   & \textit{} & ~ & R & 0.010 & {0.361} & 0.158 & 0.920 \\ \cline{3-8}
   & \textit{} & R & P & 0.007 & {0.287} & 0.096 & 0.853 \\
   & \textit{} & ~ & R & 0.009 & {0.360} & 0.156 & 0.908 \\ \hline
  \multicolumn{8}{p{2in}}{\footnotesize{1. P--prograde, R--retrograde}}\\
  \end{tabular}
  \end{table}

For S3 planetary orbits in triples, the critical semi-major axis ratio is in the range 0.287--0.361 (compared with 0.180--0.196 for S1 orbits) with wide ranges around these values depending on the parameters of the triple, and is essentially independent of the direction of motion of the outer star. The critical semi-major axis ratio is around 25\% larger for retrograde planetary orbits than for prograde orbits, significantly larger than the 8\% difference found for S1 orbits.

  \subsubsection{Comparison with observed S3 orbits}
    The smallest and largest critical semi-major axis ratios ${a_{\mathrm{io}}/{a_{2}}}$ calculated for planets in S3 orbits, based on data in \citet{wagn2016}, are 0.00061 and 0.265 respectively. Recently discovered KELT-4 Ab appears to have a very small ${a_{\mathrm{io}}/{a_{2}}}$ of 0.000132 \citep{east2016}, as does Proxima Centauri b's 3.3$\times$$10^{-6}$ \citep{escu2016}. Although the integrations found some stable bounds with semi-major axis ratios as low as 0.009 for a few stellar configurations, there were very few critical ratios lying below 0.1 and most ranged from 0.1--0.6, with a few as high as 0.9. This suggests that there may be planets at much greater distances from their outer host stars than discovered to date.

  \subsubsection{S3 orbits in triples compared with binaries}
  For S3 orbits the triple was reduced to a binary by merging the inner binary into a single star, as for the P1/P2 case, after which the integrations were repeated. The results of this approximation for the prograde stellar case are compared in Table \ref{table:19}.

  \begin{table}[htbp]
  \centering
      \caption{Difference between mean critical semi-major axis ratios in triples and binaries for S3 orbits.}
      \label{table:19}
      \begin{tabular}{l|ccc}
  \hline
  Planetary & \multicolumn{3}{c}{S3: ${a_{\mathrm{io}}/{a_{2}}}$} \\
  orbit motion& Triple & Binary & $\Delta$(\%) \\ \hline
  Prograde & 0.289 & 0.309 & 7* \\
  Retrograde & 0.361 & 0.368 & 2 \\ \hline
  $\Delta$(\%) & 25 & 19 & - \\ \hline
  \multicolumn{4}{p{2in}}{\footnotesize{* significant at the 5\% level}} \\
  \end{tabular}
    \end{table}

  The S3 orbits in a triple system have a smaller mean critical semi-major axis ratio than in a binary, since the inner binary in a triple has a larger effect on the planet than a single body of equal mass at the same distance. The average difference in S3 orbits for binaries and triples is around 5\%, which is much smaller than the average 50\% difference found for S1 orbits and more comparable with the average (absolute) 9\% difference for P1/P2 orbits. The regressions for the binary case are compared with those for the triple case in Table \ref{table:20}.

      \begin{table}[htbp]
      \centering
      \caption{Regression coefficients and model fits for S3 orbits, triples compared with binaries, for regression equations $a_{\mathrm{io}}/{a_{2}},a_{\mathrm{oi}}/{a_{2}}=C+b_{1}\mu_{1}+b_{2}\mu_{2}+b_{3}e_{1}+b_{4}e_{2}+b_{5}i_{2}+b_{6}\Omega_{2}+b_{7}\omega_{2}$.}
      \label{table:20}
      \begin{tabular}{lcrcrr} \hline
Parameter & \multicolumn{2}{c}{Triple} & \multicolumn{2}{c}{Binary} & $\Delta$Coeff. \\ \cline{2-5}
  & Coeff. & $t$ & Coeff. & $t$ & (\%) \\ \hline
\multicolumn{6}{c}{S3 orbits, prograde planet} \\ \hline
$C$ & 0.393  & 35.2  & 0.385 & 67.5 & -2 \\
$a$ & 0.000  & 20.4  & 0.000 & 26.6 & -99 \\
$\mu_{1}$ & -0.075 & -5.2  & - & - & - \\
$\mu_{2}$ & 0.037  & 11.0  & 0.062 & 28.1 & - \\
$e_{1}$ &  -0.020 & -1.9  & - & - & - \\
$e_{2}$ & -0.604 & -58.6 & -0.585 & -83.1 & -3 \\
$i_{2}$ & 0.000  & 0.0   & - & - & - \\
$\Omega_{2}$ & 0.000  & 0.3   & 0.000 & -2.8 & - \\
$\omega_{2}$ & 0.000  & 0.9   & 0.000 & -0.6 & - \\ \hline
$R^{2}$ & 0.712  &  & 0.721  &   &  \\
MAPE (\%) & 24.7 &  & 30.9  &   &  \\ \hline
\end{tabular}
\begin{tabular}{lcrcrr}
\multicolumn{6}{c}{S3 orbits, retrograde planet} \\ \hline
$C$ & 0.444  & 47.8  & 0.446 & 88.8 & 1 \\
$a$ & 0.001  & 9.1   & 0.000 & 6.1 & -100 \\
$\mu_{1}$ & 0.001  & 0.1   & - & - & - \\
$\mu_{2}$ &  0.113  & 39.0  & 0.114 & 58.3 & - \\
$e_{1}$ &  -0.011 & -1.3  & - & - & - \\
$e_{2}$ & -0.580 & -64.6 & -0.599 & -96.7 & 2 \\
$i_{2}$ & 0.000  & -0.5  & - & - & - \\
$\Omega_{2}$ & 0.000  & -3.4  & 0.000 & -1.7 & - \\
$\omega_{2}$ & 0.000  & -4.3  & 0.000 & -3.1 & - \\ \hline
$R^{2}$ &  0.790   &   & 0.810  &   &  \\
MAPE (\%) & 16.8  &   & 22.4  &   &  \\ \hline
\end{tabular}
   \end{table}

  Aside from the constant, the only orbital element of importance is $e_{2}$, the eccentricity of the outer star containing the S3 orbit. The next largest influence, that of the outer mass ratio $\mu_{2}$, is far smaller. The effect of the mass ratio $\mu_{1}$ is smaller still, unlike in the S1 case, where it was of equal quantitative importance to $e_{2}$. These coefficients are virtually identical for prograde and retrograde planetary motions as well as the direction of motion of the outer star.

  \subsection{Comparison with previous work}
The only empirical work on triples to compare with is that of \citet{verr2007}. Our prograde P1 mean critical ratio of 0.383 is 18\% smaller than the 0.466 from that study, while our P2 mean critical ratio of 2.94 is close to their 2.92. The reason for the P1 discrepancy is not clear. Although a larger parameter space and larger clouds of test particles were used in our study, the boundary of P1 orbits is always much more tenuous than that for P2 orbits, and the selection of a density cutoff to define its edge is necessarily arbitrary. So while results should be consistent within a study, a difference of this magnitude across studies is quite possible for P1 orbits.

\begin{figure}[htbp]
     \centering
     \includegraphics[width=\hsize]{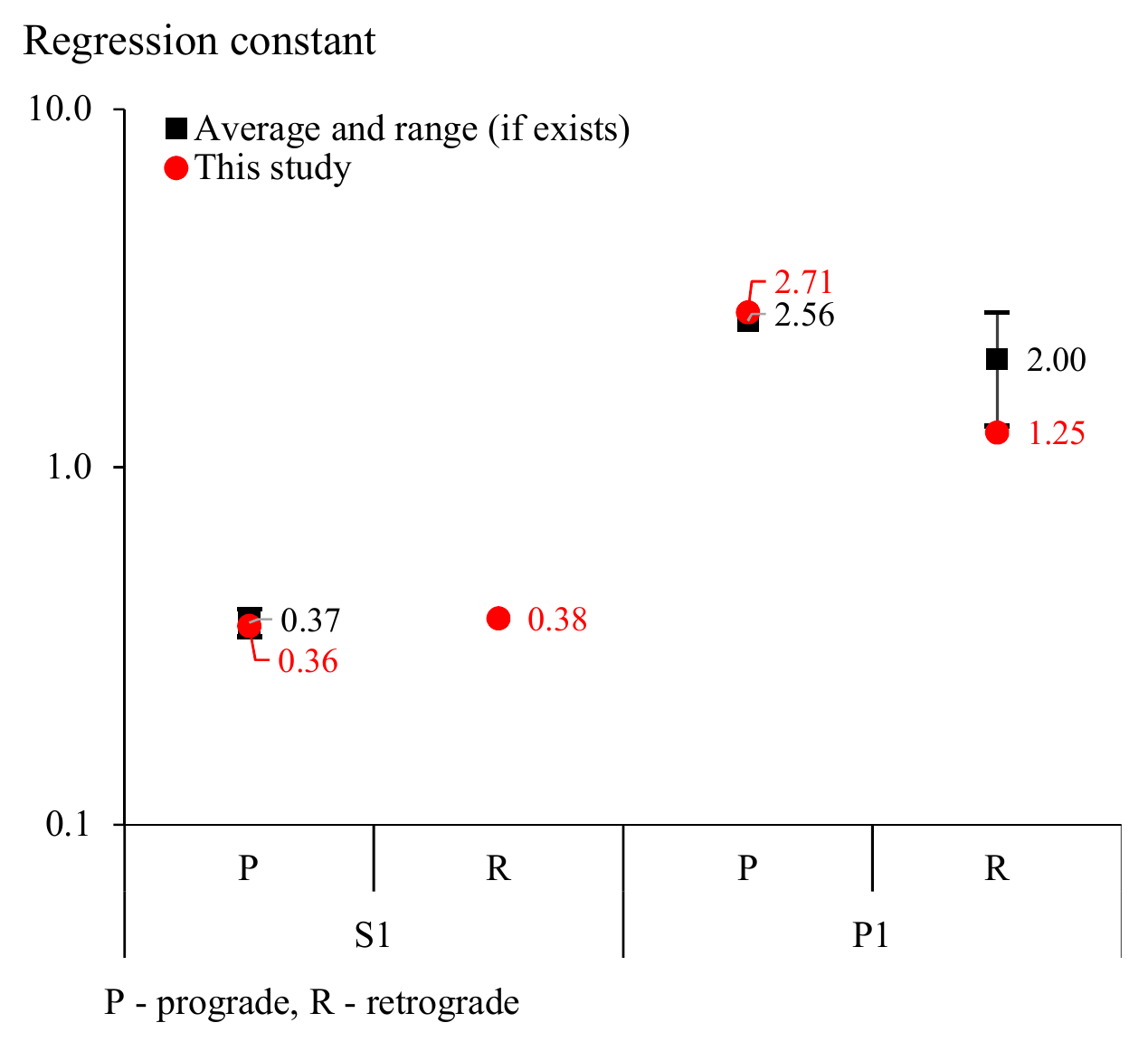}
        \caption{Comparison of results for binaries – regression constants.}
  \label{Fig10}
     \end{figure}

  A larger but still sparse area of overlap exists between our “triple-reduced-to-binary” cases and previous work on binaries. Regression constants from this study are compared with those of previous empirical studies, where provided, in Fig. \ref{Fig10}.

\begin{figure}[htbp]
     \centering
     \includegraphics[width=\hsize]{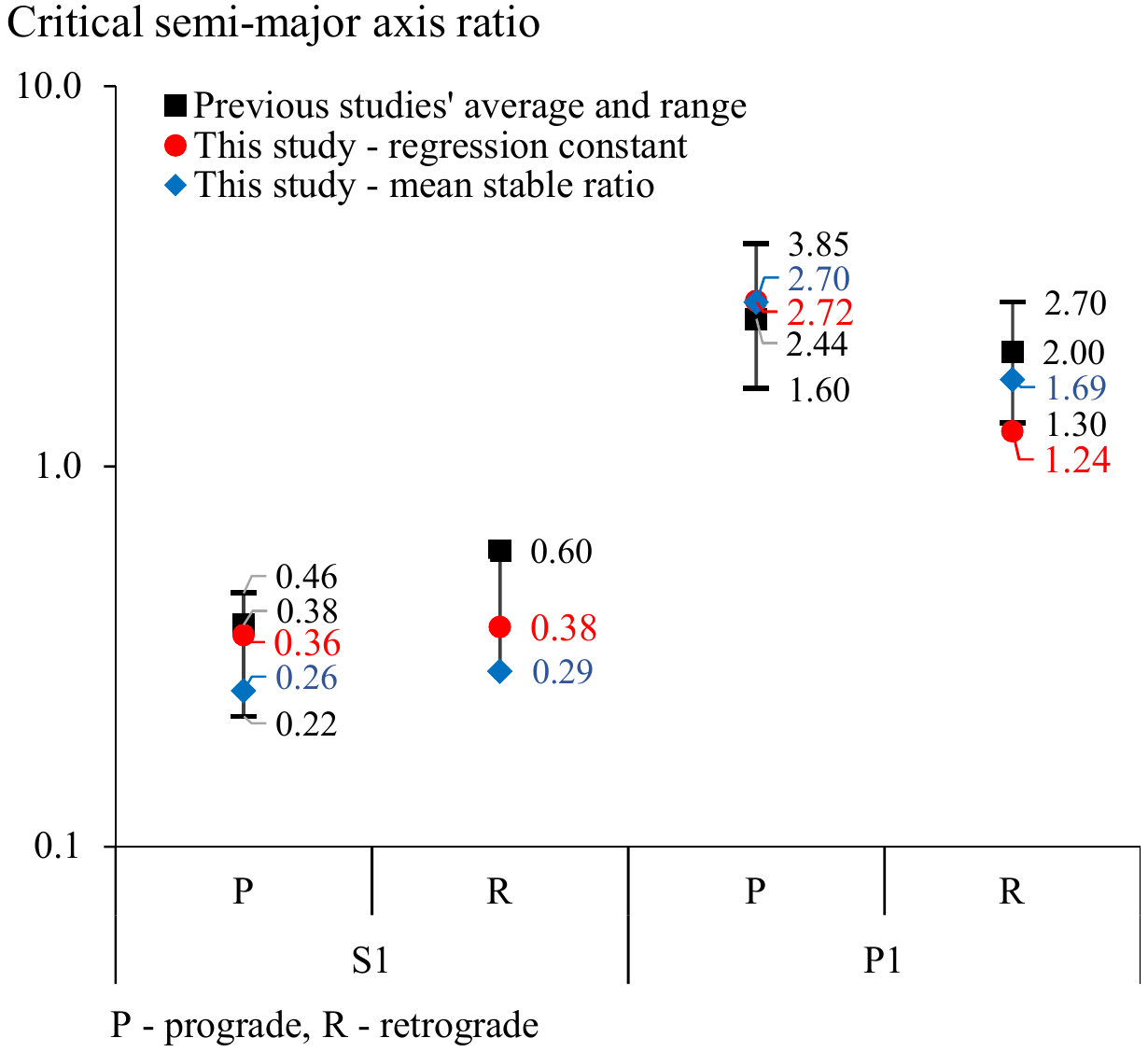}
        \caption{Comparison of results for binaries – critical semi-major axis ratios and regression constants.}
  \label{Fig11}
     \end{figure}

  For prograde S1 and P1 orbits our results are very close to those in previous work. There have been no empirical studies on S1 retrograde orbits. For P1 retrograde orbits there are two other results; ours coincides with the lower one, from \citet{dool2011}. The critical semi-major axis ratios found in previous research are shown in Fig. \ref{Fig11}, together with our ratios and regression constants for comparison.

  For binary S1 prograde planetary orbits, previous results for the critical semi-major axis ratio varied widely, ranging from 0.22--0.46 and averaging 0.38. Our corresponding mean critical ratio of 0.26 and regression constant of 0.36 are within this range. For P1 prograde orbits the range is 1.60--3.85, the study by \citet{holm1999} being the lowest, and with an average of 2.44. Our mean critical ratio and constant of 2.70 and 2.71 respectively are 11\% higher than this.

  Two empirical studies have addressed retrograde planetary orbits in binaries. The one for S1 orbits finds a critical ratio of 0.60, with our results for the mean critical ratio and constant, of 0.29 and 0.38 respectively, being around 45\% lower. The study for P1 orbits, by \citet{dool2011}, finds critical ratios in the range 1.3--2.7, comparable with our results that have a mean critical ratio of 1.69 and a regression constant of 1.25.
Despite the small sample sizes, the above results generally compare well. The one relatively large difference, for retrograde S1 orbits, suggests further investigation.

  \section{Conclusions}
  We have investigated numerically the long-term stability of planets in a large number of possible orbits around the stars in a triple system and provided a generalized mapping of the regions of stability. This was done for prograde and retrograde motion of the planets and for the outer body of the triple, as well as for highly-inclined orbits of the outer body. The construction of multiple regression equations resulted in 24 semi-empirical models, one for each type of orbital configuration.

  The greatest influences on a planet’s critical semi-major axis ratio are: for P1 and P2 orbits, the eccentricity of the outer body and, to a far lesser extent, the inner mass ratio; for S1 orbits, the inner binary’s eccentricity and inner mass ratio; and for S3 orbits, the outer body's eccentricity and outer mass ratio. Further detail can be found in \citet{buse2018}.

  We extended the number of parameters used to all relevant orbital elements of the triple’s stars and their mass ratios. We also expanded these parameters to wider ranges that will accommodate recent and possibly future observational discoveries.
The investigation of how the regions of planetary stability in triples differed from those in binaries enabled comparison with, and confirmation of, some previous studies. It also contributed some new data points.

  These generalized results can be useful in the investigation of observed systems, providing a fast method of determining their stability bounds within the large parameter space that results from observational uncertainties. The relationships expressed in the regression models can also be used to guide searches for planets in triple systems and to select candidates for surveys of triple systems. The geometry of the stable zone indicates not only where to look for planets but also the most appropriate technique to search for them.

\bibliographystyle{aa} % style aa.bst
\bibliography{TriplesPaper} % your references Yourfile.bib

\end{document}